%% file: ms_j2150_na.tex
\documentclass{nature}
\usepackage[utf8]{inputenc}
\usepackage[T1]{fontenc}
\usepackage{booktabs}
\usepackage{threeparttable}
\usepackage{multirow}
\usepackage{diagbox}
\input{preamble.tex}

\usepackage{amsmath, amssymb, amsfonts}
\makeatletter

\let\saved@includegraphics\includegraphics
\AtBeginDocument{\let\includegraphics\saved@includegraphics}
\renewenvironment*{figure}{\@float{figure}}{\end@float}
\makeatother

\makeatletter
\def\@fnsymbol#1{\ensuremath{\ifcase#1\or \dagger\or \ddagger\or
 \mathsection\or \mathparagraph\or \|\or **\or \dagger\dagger
 \or \ddagger\ddagger \else\@ctrerr\fi}}
\makeatother

\title{A 85-second X-ray quasi-periodicity after the stellar tidal disruption
by a candidate intermediate mass black hole
}

\author{Wenjie Zhang$^{1,2}$, Xinwen Shu$^{1\dag}$,  Luming~Sun$^{1\dag}$, Rong-Feng Shen$^{3}$, Liming Dou$^{4}$, Ning Jiang$^{5}$, Tinggui Wang$^{5}$ }

\begin{document}
\newcommand{\srcs}{{\rm\,J2150-0551}}
\newcommand{\src}{{\rm\,J2150-0551 }}

\newcommand{\apj}{{\it Astrophys. J.}}
\newcommand{\spie}{Proc. SPIE}
\newcommand{\pasp}{Publ. Astron. Soc. Pac.}
\newcommand{\apjs}{\it {Astrophys. J. Supp.}}
\newcommand{\araa}{{\it Annu. Rev. Astron. Astrophys.}}
\newcommand{\mnras}{{\it Mon. Not. R. Astron. Soc.}}
\newcommand{\apjl}{{\it Astrophys. J. Lett.}}
\newcommand{\aap}{{\it Astron. Astrophys.}}
\newcommand{\aj}{{\it Astron. J.}}
\newcommand{\nat}{{\it Nature}}
\newcommand{\na}{{\it Nat. Astron.}}
\newcommand{\nc}{{\it Nat. Commun.}}
\newcommand{\nar}{{\it New Astron. Rev.}}
\newcommand{\sci}{{\it Science}}
\newcommand{\aaps}{A\&AS}
\newcommand{\rmxaa}{Revista Mexicana de Astronomia y Astrofisica Serie de Conferencias}
\newcommand{\ssr}{{\it Space Sci. Rev.}}
\newcommand{\prd}{{\it Phys. Rev. D.}}

\newcommand{\gt}{$>$}
\def\ltsima{$\; \buildrel < \over \sim \;$}
\def\simlt{\lower.5ex\hbox{\ltsima}}
\def\gtsima{$\; \buildrel > \over \sim \;$}
\def\simgt{\lower.5ex\hbox{\gtsima}}
\newcommand{\msun}{{\rm\,M$_\odot$}}

\newcommand{\xmm}{{XMM-Newton} }
\newcommand{\xmmp}{{XMM-Newton}}
\newcommand{\ergs}{${\rm erg \ cm^{-2} \ s^{-1}}$ }
\newcommand{\erg}{${\rm erg \ s^{-1}}$ }
\def\uJy    {$\mu$Jy}
\def\ltsima{$\; \buildrel < \over \sim \;$}
\def\simlt{\lower.5ex\hbox{\ltsima}}
\def\gtsima{$\; \buildrel > \over \sim \;$}
\def\simgt{\lower.5ex\hbox{\gtsima}}
\newcommand{\sfr}{{\rm\,M$_\odot$\,yr$^{-1}$}}
\newcommand{\lsun}{{\rm\,L$_\odot$}}
\newcommand{\ppm}{{$\pm$}}
\newcommand{\vdag}{(v)^\dagger}

\newcommand{\chandra}{{\it Chandra} }
\newcommand{\chandrap}{{\it Chandra}}
\newcommand{\swift}{{Swift} }
 \newcommand{\swiftp}{{Swift}}
 \newcommand{\nustar}{{NuSTAR} }
 \newcommand{\nustarp}{{NuSTAR}}

\captionsetup[figure]{labelfont={bf},name={Fig.},labelsep=period}

\maketitle

\begin{affiliations}
 \item {Department of Physics, Anhui Normal University, Wuhu, Anhui 241002, China}
 \item  {National Astronomical Observatories, Chinese Academy of Sciences, Beijing 100101, China}

\item {School of Physics and Astronomy, Sun Yat-Sen University, Zhuhai 519082, China}
\item{Department of Astronomy, Guangzhou University, Guangzhou 510006, China}
\item{CAS Key Laboratory for Researches in Galaxies and Cosmology, Department of Astronomy,
University of Science and Technology of China, Hefei, Anhui 230026, China}
\item[]{$^{\dag}$\it e-mail: xwshu@ahnu.edu.cn; sunluming@ahnu.edu.cn}
\end{affiliations}

\beginbodyfigures

\begin{abstract}

It is still in dispute the existence of intermediate-mass black holes (IMBHs) with a mass of $\sim$$10^3-10^5$
solar masses (\msun), which are the missing link between stellar-mass black holes (5-50 \msun) and supermassive black holes
($10^6-10^{10}$\msun). 
The bright flares from tidal disruption events (TDEs) provide a new and direct way to probe IMBHs. 
3XMM J215022.4-055108 is a unique off-nuclear X-ray transient which can be best explained as the TDE by an IMBH in a massive star cluster, though its mass is not well determined.
Here, we report the discovery of a transient X-ray quasi-periodicity signal from 3XMM J215022.4-055108 with a period of $\sim$85 second
(at a significance of $>$3.51$\sigma$) and fractional root-mean-squared amplitude of $\sim$10\%.
Furthermore, the signal is coherent with a quality factor $\sim$16.
The significance drops to $>$3.13$\sigma$ if considering all light curves with sufficient quality for QPO search. 
 Combining with the results from X-ray continuum fittings, the detection of QPO allows for joint constraints on the black hole mass and dimensionless spin in the range $[9.9\times10^3-1.6\times10^4$\msun$]$ and $[0.26-0.36]$, respectively.
 This result supports the presence of an IMBH in an off-nuclear massive star cluster and  
 may open up the possibility of studying IMBHs through X-ray timing of TDEs.
\end{abstract}

Black holes with masses in the range of $10^3-10^{5}$ \msun~are particularly important.
By filling the mass gap between stellar mass black holes (sBHs) with $M_{\rm BH}$$<$$100$\msun~and
supermassive black holes ($M_{\rm BH}$$\gtrsim$$10^6$\msun) at the centre of most galaxies, these intermediate-mass black holes
(IMBHs) are potential analogues of the seeds of supermassive BHs\cite{Volonteri2010,Greene2012}.
In current models of galaxy evolution in a hierarchical cosmology, IMBHs are expected to exist in smaller stellar systems
such as dense star clusters\cite{Portegies2002} and dwarf galaxies\cite{Volonteri2009}, which may have not enough time to be fully grown.
The mass function and accreting properties of IMBHs in the nearby universe are thus crucial in discriminating
different models for seed BHs\citep{Volonteri2008, Greene2020}.

However, the search for IMBHs turns out to be a nontrivial task.
The gravitational influence radius of IMBHs is usually too small to be resolved for
direct mass measurements using star or gas dynamics, except for a handful of the nearest ($\lesssim$1 megaparsec) star clusters and galaxies \cite{Greene2020}.
While they can reveal themselves as active galactic nuclei (AGNs) in nearby
dwarf galaxies,
very few have been reliably identified to have black holes in the mass range below $10^5$\msun\citep{Dong2007, Reines2013, Baldassare2015, Woo2019}.
This raises a question as to whether there is a turn over of the mass function toward the lower end
or there exists a large population yet to be discovered\citep{Chilingarian2018, Greene2020}.
Although globular clusters and off-nuclear ultraluminous X-ray sources (ULXs, $L_{X} > 10^{39}$ erg s$^{-1}$)
have been proposed to harbour less massive IMBHs, observational properties for their existence are still debated\citep{Tremou2018, Kaaret2017}.

3XMM J215022.4--055108 (hereafter, J2150-0551) is a luminous transient X-ray source \citep{Lin2018} located in a large lenticular
galaxy at $z=0.055$, with a positional offset to the galactic center by $\sim$12 kpc.
The X-ray source was first detected and peaked in 2006 with a luminosity of $\sim$10$^{43}$ erg s$^{-1}$.
It then decayed slowly, following approximately a $t^{-5/3}$ decline law for over ten years.
The X-ray spectra are supersoft, which lack emission at energies above $\sim$3 keV and can be described
by a soft thermal component, presumably arising from an accretion disk.
The outburst X-ray emission with a high peak luminosity, the supersoft spectral state and
the power-law decay of the luminosity evolution with an index of $t^{-5/3}$,
strongly suggest that \src is likely associated with a tidal disruption event (TDE), in which a strayed star was tidally disrupted by a massive
BH \citep{Rees1988}.
The BH mass was estimated to be $\lesssim$$5\times10^{4}$\msun~based on the fit to the X-ray spectra with various
disk models\citep{Lin2018,Lin2020,Wen2021}, indicating the presence of an IMBH in \srcs. 
The IMBH TDE scenario is further supported by the detailed modelling of the X-ray light curve \citep{Chen2018},
which can be described by the tidal disruption of a main-sequence star of $M_{\ast}\simeq0.33$\msun~by an IMBH with $M_{\rm BH}\sim7\times10^4$\msun. 
At the explosion site of \srcs,  a massive cluster of stellar mass $\sim10^7$\msun~was inferred \citep{Lin2020}.
IMBH TDE candidates have been reported in other optical and X-ray transients \citep{Krolik2011, Donato2014, Perley2019, Angus2022, Zhang2022},
though the luminosity evolution is much faster than that observed in \srcs.
Therefore, TDEs offer a unique opportunity to find and study IMBHs, especially in those otherwise quiescent dwarf galaxies and massive star clusters.

The outburst X-ray emission in J2150-0551 may exhibit quasi-periodic oscillation (QPO) as found in accreting sBHs\citep{Motta2016, Zhang2023}
as well as supermassive black holes in the centres of galaxies \citep{Lin2013, Pasham2019}.
To check for the presence of X-ray QPO in \srcs,
we produced the light curves over the energy band 0.2--2 keV for all three observations of \src
from the X-ray Multi-Mirror Mission (XMM-Newton).
However, a significant fraction of the \xmm data were severely affected by background flares.
We carefully removed the high background flux epochs from our analysis, resulting in several
light curve segments in each observation.
Following the general procedures of timing analysis\cite{Klis1989, Vaughan2003}, we then Fourier transformed them
into power density spectra (PDS) in order to search for a potential QPO signal (see "Method": Power spectral analysis).
The PDS of one light curve segment in second \xmm observation (hereafter XMM2) displays
an apparent QPO component at a frequency of $11.88 \pm 0.38$ mHz (Figure 1a and 1b), corresponding to a period of 84 s.
A more precise estimation of the period can be made by employing the epoch folding technique \citep{Leahy1983},
which is found to be 85 s.
The power value at 11.88 mHz becomes stronger if we combine the data acquired by all three detectors (PN, MOS1 and MOS2)
for the \xmm observation.
 The QPO concentrates in just a few frequency bins of the unbinned PDS, which makes it highly coherent, with a quality factor $\nu/\delta\nu\sim16$, where $\nu$ is the centroid frequency and $\delta\nu$ is the FWHM of the Lorentzian model to fit the QPO (Extended Data Fig. 5).
The fractional root-mean-squared (rms) variability amplitude in the QPO calculated from the folded light curve (Figure 1c) is 10.1$\pm$2.3\%.


As shown in Figure 1b, the PDS is dominated by white noise, and the frequency-dependent red noise is weak (see "Method": Noise properties in PDS).
Under the white noise hypothesis, we utilized the $\chi^2$-statistics to analytically estimate the global false alarm probability of the QPO signal at 11.88 mHz,
which is $P_{\rm false}\sim$$ 5.6\times10^{-5}$, equivalent to a global significance of 4.03$\sigma$ for a normal distribution (see "Method": QPO significance).
In order to further quantify the statistical significance of the QPO feature
by properly accounting for the underlying red noise,
we conducted rigorous Monte-Carlo simulations. 
These were done by comparing the spread of a series of simulated power spectra to the observed one. 
The Monte-Carlo approach we adopted
is similar to that outlined in ref. \cite{Barret2012, Reis2012, Lin2013, Pasham2019},
which is essentially based
on the well-established procedures in X-ray timing studies of accreting compact objects\cite{Lewin1988, Uttley2002}.
In this case, we obtained a conservative lower limit of false-alarm probability $P_{\rm false}\sim  4.5\times10^{-4}$ for the QPO at 11.88 mHz (or a significance level of 3.51$\sigma$ assuming a normal distribution).




It should be noted that the QPO was present only for a short duration in XMM2,
as the signal comes down if the whole light curve was used,
indicating that the QPO might be a transient phenomenon.
We considered the $\sim$8 ks light curve segment in the first XMM-Newton observation (XMM1) as an independent search, and we excluded the third XMM-Newton observation (XMM3) from further analysis as the count rate is too low to yield a positive
detection for QPO (see "Method": Power spectral analysis).
Then, among a total exposure time of $\sim$58 ks, the QPO appeared in the first segment with $\sim$17 ks length
between two high background intervals (Figure 1a). 
This is equivalent to detecting QPO in one out of $\sim$66 ks/17 ks $\approx$ 3.88 searches.
Considering 
the searches in the light curves obtained in XMM1 and XMM2 gives
a conservative lower limit of false-alarm probability $P_{\rm false}\sim$ 3.88 $\times$ (  4.5 $\times$ 10$^{-4}$) $\sim$   1.75$\times10^{-3}$ (or   3.13$\sigma$ assuming a normal distribution),  
making the QPO detection still statistically significant. 
The QPO detection is likely reliable since the signal becomes stronger when the data from the PN and MOS detectors are combined.
Furthermore, we performed searches for QPO in the PDS of the background sky and several bright X-ray sources
in the same field of view and epoch as \srcs, and no signal similar to that at 11.88 mHz was found.
Based on these tests, we ruled out an instrumental or a background as the origin for this QPO signal (see "Method": Extended Data Fig. 10).
Note that some sBHs and AGNs have also shown transient QPOs in their X-ray light curves,
though at much different frequencies
\citep{Belloni2012, Remillard2006, Gierlinski2008, Pan2016, Zhang2017}, suggesting that the temporal appearance of QPO may be common among BH accreting systems. 

Having established that the QPO in \src is statistically significant,
we now place its constraint on the origin of the X-ray emission and
the central engine of \srcs.
The discovery of 11.88 mHz QPO with a significance $>$3.5$\sigma$ provides strong evidence that the central engine is a compact object, which could be either a neutron star or an accreting black hole.  mHz QPOs have been detected in a few neutron-star low-mass X-ray binaries\citep{Revnivtsev2001, Tse2021} , which are generally explained as thermonuclear burning on the neutron star surface. Therefore, these QPOs are transient behaviours that occur only in a narrow range of X-ray luminosity
before the thermonuclear burst and then disappear afterwards.
This contradicts the 11.88 mHz QPO in \srcs, as it is in the decay phase where the luminosity has dropped by an order of magnitude over $>$5 years since the peak.
In addition, mHz QPO in neutron stars has a rms amplitude of order 1\%\citep{Revnivtsev2001}, which is much smaller than that observed in \srcs ($\sim$10\%).
Also, based on the detection of extended emission in its optical counterpart, 
previous work \cite{Lin2020} ruled out the explanation of Galactic cooling neutron star in a low-mass X-ray binary for \srcs.
Although extreme ULXs with X-ray luminosity $\gtrsim$ 10$^{41}$ erg s$^{-1}$ could be powered by an accreting pulsar \citep{Israel2017},
such a scenario is unlikely to account for \src
due to the TDE-like long-term X-ray luminosity evolution and the very soft X-ray spectrum \citep{Lin2018, Chen2018}.
In addition, the QPO spans multiple frequency bins in the unbinned PDS ("Method": Extended Data Fig. 5), which suggests that
the signal is not strictly periodic as expected in pulsar's emission.
We conclude that models involving a neutron star is challenging to explain the multi-wavelength properties of \srcs.


There are a few accreting sBHs that have shown
QPOs with frequency in the mHz range.
However, most of these sources are believed to have
high inclinations of processing accretion disk with respect to the line of sight ($\sim60^{\circ}-70^{\circ}$)\cite{Cheng2019}.
Given the high average X-ray luminosity ($>$$10^{42}$\erg) and supersoft X-ray spectra with little absorption
intrinsic to the X-ray source\cite{Lin2018, Lin2020}, \src is unlikely to be a high inclination sBH system.
Furthermore, these sources often show the red noise PDS below $\sim3-5$ mHz \citep{Pasham2013, Pahari2017}.
These properties distinguish them from the \srcs's 11.88 mHz QPO, which appears with negligible red noise in the PDS.
On the other hand, if isotropically emitted, the maximum luminosity of \src requires that it is radiating at
$\simgt10^{4}$ times the Eddington limit for a sBH ($M_{\rm BH}\sim10$\msun), challenging standard
models of black hole accretion.
Therefore, if the mHz QPO of \src was due to a process similar to that operating in sBHs,
the underlying physical mechanism would be extreme and have never been seen before.

If the centroid frequency of the observed 11.88 mHz QPO was set by the Keplerian frequency of the innermost stable circular orbit
(ISCO, the highest characteristic variability frequency),
it would imply a black hole mass of $\sim$$1.8\times10^{5}$\msun~or $2\times10^{6}$\msun,
for a non-rotating and maximally rotating black hole, respectively.
QPOs with frequencies at a few hundred hertz (HFQPOs) have been occasionally seen in several sBHs
with dynamic mass constraints. They sometimes occur in pairs with a constant frequency ratio of 3:2.
Unlike the low-frequency QPOs at $\sim0.1-20$ Hz, these frequencies ($f_{\rm HFQPO}$) appear stable and are
considered a fundamental property that might be linked to the BH mass and spin.
A tentative inverse linear relation between $f_{\rm HFQPO}$ and BH mass was suggested based
on observations of three sBHs \cite{Remillard2006}, which can be extended to more massive BHs \citep{Zhou2015, Pan2016, Song2020}.
Although HFQPOs were mostly found in hard X-rays ($>$2 keV) and in the steep power-law state of sBHs,
if assuming that the 11.88 mHz QPO of \src
corresponds to the stronger $2\times f_{\rm 0}$ of the 3:2 frequency pair (where $f_{\rm 0}$
is the fundamental frequency) and follows the frequency-mass relation proposed for HFQPOs\citep{Remillard2006, Aschenbach2004},
the BH mass in \src can be inferred in between 3.6$\times10^4$\msun~and 1.7$\times10^5$\msun~(Figure 2),
depending on the extent of the spin constraints. 

By modelling the X-ray energy spectra with a standard thin or a slim disk model,
previous works\citep{Lin2018, Wen2021} suggested that \src hosts an IMBH with mass in the range of $10^{4}-10^5$\msun.
These measurements have large uncertainties owing to the degeneracy of model parameters
between mass, spin and accretion rate.
It is worth investigating whether the BH mass and spin can
be better constrained by the joint analysis of QPO frequency and the X-ray spectral fitting results.
We started our analysis by associating the QPO frequency with the three fundamental frequencies of a test particle moving at the ISCO, namely 
the Keplerian orbital frequency, the vertical epicyclic frequency and the Lense-Thirring (LT) precession frequency.
The resulting BH spin versus mass contours are shown in Figure 3.
In addition,
we also performed independent spectral fittings to the data taken in the 2009 XMM-Newton observations (XMM2) in which the QPO is detected, adopting the same model and fitting procedures as described in ref.\citep{Wen2021} (see "Method": X-ray spectral analysis and Extended Data Fig. 11).
This will help reduce the uncertainty in measuring mass and spin due to accretion rate variations.
The 68\% and 90\% confidence contours of BH spin versus mass are shown in Figure 3
with the grey dotted and solid lines, respectively.
Interestingly, we find that the contours derived from the X-ray spectral fittings intersect with that constrained by QPO if its frequency is related to the LT precession frequency.
At the 90\% confidence level, the BH mass can be constrained in the range $[9.9\times10^3-1.6\times10^4$\msun$]$,
while the BH spin is in the range of $[0.26-0.36]$.
Placing the QPO at a larger radius,
e.g., 10 $R_{\rm g}$ where $R_{\rm g}$(=$GM_{\rm BH}/c^2$) is the gravitational radius,
would shift the limit on BH spin to higher values,
but the contours start to exclude mutually. 
Therefore, the joint observational constraints suggest that the spin cannot exceed 0.4 at a 90\% confidence level, indicating that the BH may have a moderate spin.

The above analysis also indicates that the QPO may originate from a radiating material close to ISCO.
Such a QPO could be driven by the LT precession of tilted, narrow inner sub-disks.
This is possible as in a TDE, the stellar debris' orbit and the BH spin axis will be misaligned, and the disk material is expected to undergo LT precession \citep{Stone2012, Franchini2016}.
The GRMHD simulations performed by ref. \citep{Kaaz2023} have shown that LT precession causes several torn sub-disks, and a precessing inner sub-disk
could produce a QPO with a frequency of $\sim$3 Hz for a 10\msun~black hole \citep{Musoke2023}.
Scaling the black hole mass to that of J2150-0551 would predict an LT precession-driven QPO at several mHz, which is not at odds with the 11.88 mHz observed in \srcs.
In addition, the simulations also suggest that if the radiating region producing the QPO is made up of many stochastically distributed
tearing rings \citep{Musoke2023}, it will start de-coupling after a few LT precession cycles, which may explain
why the J2150-0551's QPO is a transient phenomenon.

 A handful of optical and X-ray detected transients in dwarf galaxies had been proposed to be powered by IMBH TDEs\citep{Donato2014, He2021, Angus2022},
 but their X-ray emission is either too weak or lacking sufficient signal-to-noise ratios for a QPO search.
 In addition to \srcs, the only other two TDEs exhibiting an X-ray QPO are Swift J1644+57\citep{Reis2012}
 and ASASSN-14li\citep{Pasham2019}, both
 occurred in the centre of a galaxy.
 Swift J1644+57 is a peculiar TDE with a relativistic jet that dominates the electromagnetic output, including the X-rays\citep{Reis2012}.
 It has been proposed\cite{Krolik2011} that Swift J1644+57 might be powered by the tidal disruption of a white dwarf by an IMBH ($M_{\rm BH}\lesssim10^5$\msun).
ASASSN-14li is a more typical TDE with supersoft X-ray spectra that can be described entirely by thermal emission\citep{vanVelzen2016}.
 The X-ray timing and spectral properties, i.e., white noise PDS, QPO rms amplitude, and supersoft X-ray spectrum,
 make \src similar to the thermal TDE ASASSN-14li\cite{Pasham2019}.  
However, the BH mass in ASASSN-14li was estimated in the range $10^6-10^7$\msun~using
the measurements of the host galaxy's properties\cite{Pasham2019}.
Since \srcs's periodicity was comparable to that of ASASSN-14li ($f_{\rm QPO}=7.65$ mHz),
it would imply that either the BH mass in ASASSN-14li was overestimated or ASASSN-14li's
QPO may correspond to a different disk oscillation mode, which may not obey the frequency-BH mass relation found in sBHs (Figure 2).
In fact, \src is hosted in a stellar cluster with mass of only $\sim10^7$\msun (ref. \citep{Lin2020}), which does not support a central black hole significantly more massive than $\sim10^5$\msun (ref. \cite{Mezcua2017}),
which is in line with the results from the above joint constraints
with the QPO frequency and X-ray spectral fittings. 


 In summary, we present evidence for an 85-second X-ray QPO from \srcs,
a candidate of otherwise dormant IMBH residing in a massive star cluster at a projected distance of 12.5 kpc from the centre
of a lenticular galaxy\cite{Lin2018}.
 This is the first time such a QPO has been found after the stellar tidal disruption by an off-centre IMBH.
 Although a specific physical mechanism to produce the QPO remains unknown,
our discovery suggests that
it could be related to the LT precession frequency,
and similar signals can be used to confirm the presence of IMBHs
and potentially constrain the spin if the BH mass is well determined using other means.
Given an estimated volumetric rate of
$\sim$$10^{-8}$Mpc$^{-3}$yr$^{-1}$ for all off-centre X-ray transients like \src  (ref.\citep{Lin2018}), 
the forthcoming 
time-domain X-ray surveys with extended Roentgen
Survey with an Imaging Telescope Array\citep{Predehl2020} and Einstein Probe\citep{Yuan2022} can find one such system per year.
Since an optical flare was also detected in \srcs, this estimate would be much improved in synergy with the optical discoveries by Large Synoptic Survey Telescope\citep{Bricman2020}.
Sensitive X-ray follow-ups can then open up systematic studies of QPO, with which to probe the physical properties of accreting IMBHs.

\newpage
\beginedfigures
\begin{figure}[ht]
\centering
\includegraphics[width=\linewidth]{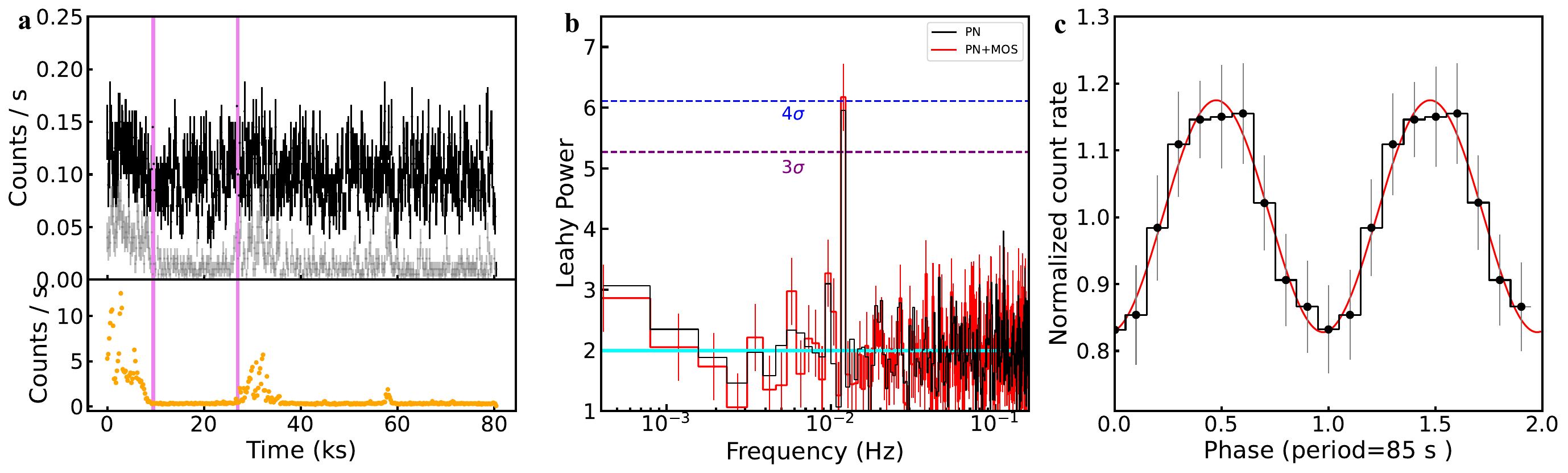}
\caption{\textbf{Figure 1: The PN light curve from the XMM2 observation, in which the QPO is detected.}
{\bf a:} We present the 0.2-2 keV light curve of source (black) and background (gray).
For a comparison, we also present the background light curve (orange) in the 10-12 keV to demonstrate the
effect of high-energy particles.
{\bf b:} Leahy-normalized PDS of \src in the 0.2-2 keV extracted from the light curve segment between two violet vertical lines on the left.
Note that the PDS was computed using the light curve encompassing both source
and background counts.
The PDS was binned to have a frequency resolution of 0.76 mHz ( "Method": Power spectral analysis). A strong peak appears at $11.88\pm0.38$ mHz
($\approx $84-second), indicating the presence of a QPO component.
The QPO frequency is defined as the centroid value of the peak frequency bin, and the error is the half of the bin width.
The red line represents the PDS of the PN+MOS data, while the black line is for the PN PDS only.
The purple and blue dashed line represent
 the $3\sigma$ and $4\sigma$ white-noise statistical thresholds
derived using the $\chi^2$ distribution, respectively.
The cyan horizontal line shows the white noise (the value is 2 in Leahy-normalized power spectrum).
{\bf c:} Folded light curve in the 0.2-2 keV with a period of 85 second,
obtained with the epoch folding technique. 
The red solid curve shows the best-fit sinusoid to the data.
The error bars represent 1$\sigma$ uncertainties.
}
\end{figure}

\begin{figure}[ht]
\centering
\includegraphics[width=\linewidth]{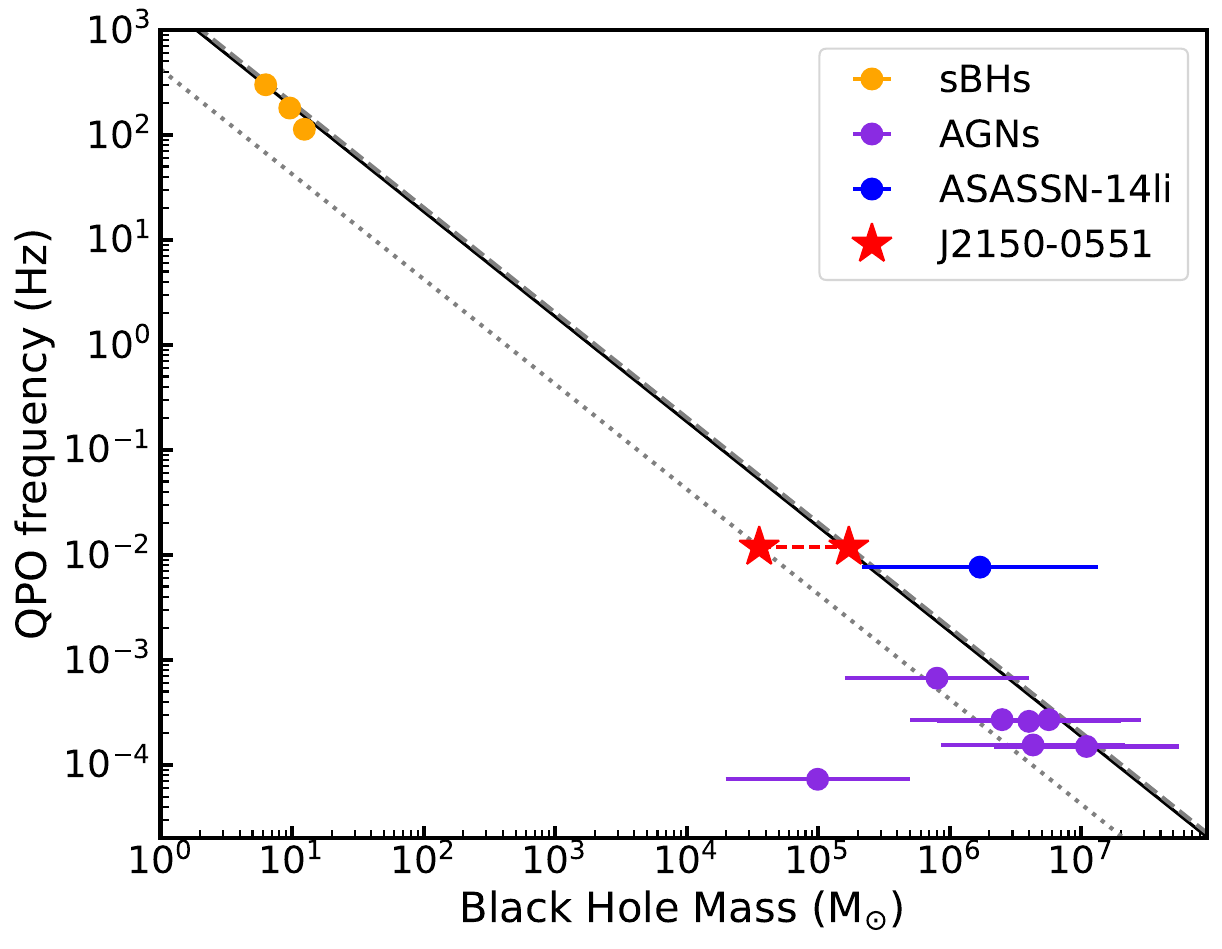}
\caption{\textbf{Figure 2: Relation between QPO frequency and BH mass.}
The black solid line is the extrapolation of the relation between QPO frequency and BH mass for sBHs derived in ref.\citep{Remillard2006},
assuming that the QPO corresponds to the 2$\times f_{0}$ for the 3:2 harmonic peaks.
The dashed and dotted lines represent the relation derived from the model of 3:2 resonance  \citep{Aschenbach2004} with the spin parameter $a=0.996$ and $a=0$, respectively.
The HFQPOs of sBHs with dynamic mass measurements are shown in filled orange circles \citep{Zhou2015}.
The data for TDE (ASASSN-14li) and AGNs hosting supermassive BHs are taken from ref.\citep{Pasham2019,Song2020}.
The \srcs's BH mass range constrained by the 11.88 mHz QPO is shown by red stars.
For ASASSN-14li and AGNs, the errors on BH masses are reported to take into account the 1$\sigma$ statistical errors and typical errors of $\sim$0.5 dex of the scaling relation
between the BH mass and host properties.
}
\end{figure}

\begin{figure}[ht]
\centering
\includegraphics[width=\linewidth]{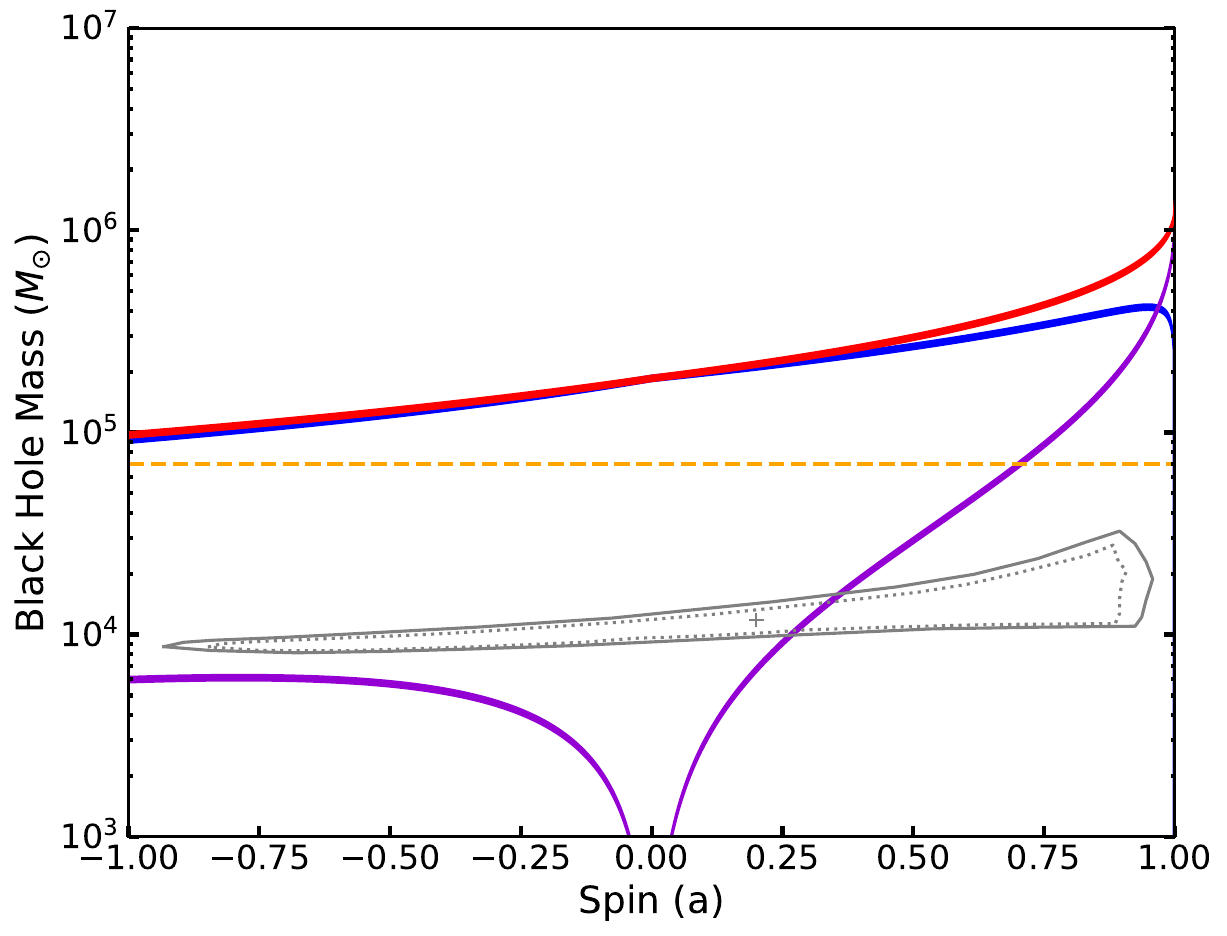}
\caption{\textbf{Figure 3: Relation between BH dimensionless spin and mass parameter.} Spin-versus-mass contours determined by assuming that the 11.88-mHz QPO is associated with any of three particle frequencies—Keplerian frequency (red), vertical epicyclic frequency (blue), and Lense-Thirring precession (violet)—at the ISCO.
The thickness of these curves reflects the width of QPO signal in the PDS, which is
0.38 mHz for \srcs.
For comparison, we also plot the 68\% and 90\% confidence contours of BH dimensionless spin versus mass, which are shown with gray dotted and solid lines, respectively.
The results are obtained by fitting the X-ray spectrum from the XMM2 observation
with a relativistic slim disk model ("Method": X-ray spectral analysis).
The orange dashed line represents the BH mass obtained through the X-ray long-term light curve \cite{Chen2018}, 
which assumes the brightest luminosity observed in XMM1 is the Eddington luminosity. 
This could be considered as an upper limit if the source were caught in the super-Eddington accretion phase.
These contours can mutually overlap only in the case assuming that the QPO frequency is associated
with the Lense-Thirring precession frequency. }

\end{figure}


\newpage
\clearpage
\begin{methods}

{\bf 1. Observation and data reduction.}
\src has been observed by three X-ray imaging telescopes, including XMM-Newton, Chandra, and Swift.
Based on these observations, ref.\citep{Lin2018, Lin2020} found that the source exhibited a prolonged
X-ray outburst, followed by a long decay in the X-ray luminosity over a decade, as shown in the Extended Data Fig. 1. The Chandra and Swift observations have too few counts for meaningful
timing analysis, thus we considered only \xmm data sets for further analysis.
XMM-Newton carries two sets of X-ray detectors, including three European Photon Imaging Cameras, namely PN, MOS1 and MOS2.
\src has three publicly available XMM-Newton observations, and we reprocessed them taken from the three cameras using Science Analysis Software version 17.0.0, with the latest calibration files. We extracted the event files from a circular region with a 35$^{\prime \prime}$ radius centred on the source position (RA: 21$^{\rm h}$50$^{\rm m}$22$^{s}$.4, Dec.: -05$^{\circ}$51$^{\prime}$08$^{\prime \prime}$, ref.\citep{Lin2018}). 
  Source spectra were extracted from this circular region, and background spectra were made from source-free areas on the same chip using four circular regions identical to the source region.
 Because \srcs's energy spectrum is extremely soft, with $>$94\% of photons in the 0.2--2 keV,
we extracted single-pixel events for both the PN and MOS, as well as single
plus double-pixel events (PATTERN$\leq 4$ for PN and PATTERN$\leq 12$ for MOS) 
in the energy range of 0.2-2 keV for light curve analysis.
 
The epochs of high background events were examined using the light curves in the energy band 10--12 keV.
We adopted a count-rate filtering criterion of 0.5 counts s$^{-1}$ for PN and
0.4 counts s$^{-1}$ for MOS to remove 
the time intervals that are affected by the high particle background, 
which resulted in several light curve segments with good time intervals (GTIs).  
The violet-shaded regions in the Extended Data Fig. 2 mark the epochs of these GTIs.
A summary of these observations can be found in Extended Data Table 1.
We also assessed the extent of photon pile-up using the SAS task {\tt epatplot} and found that such
an effect is negligible.


 {\bf \noindent 2. Power spectral analysis.}
 %
In order to improve the detection sensitivity of a QPO signal in the PDS, we combined the data from the three \xmm detectors, i.e., PN, MOS1 and MOS2.
 As the time resolution of PN (73.4 ms) and MOS (2.6 s) observations is different, we used
  a uniform time bin size of 2.6 s to combine PN, MOS1 and MOS2 light curves.
As we mentioned above, the \xmm data were broken into several GTIs after screening the high background epochs,
resulting in a total of five usable light curve segments (Extended Data Table 1).
We then Fourier transformed these light curve segments into individual PDS, which was normalized by Leahy method\cite{Leahy1983} to have a mean
Poisson noise level of 2.
 This is ensured as the PDS was
computed using the light curve encompassing both source and background counts.
All the PDS were initially rebinned to have a frequency resolution of $\sim$$0.8$ mHz,
similar to that chosen for the PDS of TDE ASASSN-14li \citep{Pasham2019}.

As shown in Extended data Fig. 3, the Leahy-normalized PDS reveals an apparent QPO component at a
frequency of $11.88 \pm 0.38$ mHz in one of the light curve segments (GTI2 in XMM2).
In order to further investigate whether there is any weak QPO signal in the PDS of other segments,
we restricted the three individual light curve segments in XMM2 to have the same exposure length of 17 ks, and then combined them to obtain an averaged PDS.
We uniformly chose
the initial 17 ks in each light curve segment, which will help to avoid
the effect of arbitrarily selected light curves that may
cause false signals in the PDS.
Since GTI5 in XMM3 has a much lower count rate by a factor of $\sim$5 ($0.02\pm0.01$ cts s$^{-1}$), it was excluded from the PDS stacking to avoid introducing too much noise in the PDS. 
We found that the power value at 11.88 mHz comes down (Extended Data Fig. 4), probably because the QPO signal is not detectable in the other light curve segments.
To account for the data obtained in XMM1, we cut the three individual light curve segments in XMM2 to have the same exposure length of 8 ks, and then combined them to obtain an averaged PDS.
Note that we split each light curve segment (GTI2, GTI3, GTI4) in XMM2 into two 8
ks segments, as they have enough exposure length ($>$17 ks).
We did not find any QPO feature in the averaged PDS, with the power values consistent with the noise level of $\sim$2.
These suggest that the QPO at 11.88 mHz was a transient phenomenon,
as otherwise, the signal would get stronger in the averaged PDS.

 For a sanity check, we investigated the QPO signal in the unbinned PDS of GTI2 (with a minimum frequency resolution of 1/17000s=$5.88\times10^{-5}$ Hz),
and found that the signal is distributed over several frequency bins (Extended Data Fig. 5).
To better understand the spread of the QPO signal,
we modelled the entire \texttt{unbinned} PDS (including the QPO) covering a frequency range between $5.88\times10^{-5}$ Hz and 0.19 Hz (1/(2$\times$2.6s)), with a model consisting of a powerlaw+constant plus a Lorentzian model.
The powerlaw+constant model accounts for the continuum, while the Lorentzian model fits the QPO signal.
We found that the addition of the Lorentzian model is statistically significant:
the $\chi^{2}$ decreased by 22.1 for three extra free parameters,
corresponding to a significance level of 99.991\% according to the F-test.
To verify the result, we also fitted the PDS after averaging the \texttt{unbinned} PDS by a factor of 2, and found that adding the Lorentzian model can also improve the fitting result.
In both cases, the resulting FWHM of the Lorentzian model is $\sim$7.61$\times10^{-4}$ Hz, which is $\sim$13 times the minimum frequency resolution of the {unbinned} PDS.
Therefore, to better illustrate the QPO feature, we will show the PDS averaged in frequency by a factor of 13 (i.e., a frequency resolution of 0.76 mHz) throughout the paper.

In addition, we also computed the PDS for the PN and MOS data separately.
We did find the same QPO signal in the PN light curve alone, though the strength is
slightly weaker (Figure 1).
The QPO signal was not detected in either individual or
combined EPIC MOS data, possibly due to the relatively low
count rates in the individual MOS light curves, which are a
factor of $\sim$3.5 less than those in PN.
However, the fact that the signal becomes stronger when combining the data from the PN and MOS detectors validates the QPO detection.
We will perform detailed Monte Carlo simulations to quantify the
significance level of the QPO in the next Section.

For the light curve segment (GTI2) in which the QPO is detected,
we tested that the signal persists throughout the entirety of GTI2, and
is not confined to specific segments of it.
We then attempted to ascertain the period of QPO by employing the epoch-folding technique \citep{Leahy1983}, and found it is 85 seconds.
This is very consistent with the period of 84 seconds derived from the centre frequency of
the QPO (11.88 mHz).
To be precise, we will refer to the period of 85 seconds for the QPO throughout the paper.
In combination with the mean source count rate of 
 0.15 cts s$^{-1}$ (PN+MOS),
we estimated the fractional rms amplitude of the QPO from the PDS\cite{Pasham2021}
as $\sqrt{P_{\nu}\times\delta\nu/S}=\sqrt{ 4.1\times7.6\times10^{-4}/ 0.15}= 14.4\pm 3.6\%$,
where $P_{\nu}$ are the power values within the QPO width ($\delta\nu$)
and $S$ is the source mean count rate.
The fractional rms amplitude is consistent (within errors) with the measurement from the folded light curve of $ 10.1\pm 2.3$\%.


{\bf \noindent 3. QPO significance.}
{\bf \noindent Noise properties in PDS: }
The statistical significance of the QPO at 11.88 mHz depends on the underlying noise distribution in the PDS.
Fig. 1 shows that the PDS appears to be flat between a few mHz and 0.19 Hz (except for the frequency where the QPO is located),
i.e., the noise powers are independent of frequency, as expected for white noise.
A statistical test for white noise examines whether the noise powers are $\chi^{2}$ distributed\cite{Klis1989}.
We removed the three frequency bins centered at 11.88 mHz, and generated a cumulative distribution function (CDF) of the powers at all other frequencies.
We then compared it with a $\chi^{2}$ distribution (with $2\times13$ degrees of freedom, d.o.f) scaled by a factor of 1/13, where the number 13 is
the factor by which the PDS was rebinned (Extended Data Fig. 7 (a)).
In addition, we also analyzed the probability density distribution of these powers and compared it with the expected values from a $\chi^{2}$ distribution.
The result is shown in the Extended Data Fig. 7 (b). Both tests show consistent results that the powers of the noise can be described with $\chi^{2}$ distribution, i.e.,
dominated by white noise.

 For more stringent tests on the noise properties,
we further performed the Kolmogorov-Smirnov (K-S) and Anderson-Darling goodness-of-fit tests, under the null hypothesis that the noise powers between 7.65$\times10^{-4}$ and 0.1 Hz (excluding the bins including the QPO feature) are $\chi^{2}$ distributed with $2\times13$ d.o.f scaled by a factor of 1/13.
 Note that we choose an upper limit of 0.1 Hz to ensure the sampling is not biased by higher (>0.1 Hz) frequencies close to the edge of PDS.
In other words, the null hypothesis to be tested is that the underlying noise is white.
The better the $\chi^{2}$ distribution fits the data, the smaller these statistics will be.

Following the procedures outlined in ref. \citep{Pasham2019, Pasham2021}, we computed the K-S test statistic using the CDF, which can be used to reject the null hypothesis.
First, we randomly drew the same number (=130) of elements as the observed noise powers from a $\chi^{2}$ distribution with $2\times13$ d.o.f. Then, we evaluated its CDF and scaled it by a factor of 1/13, just like the real data. Finally, we compute the K-S test statistic and store its value. We repeated this process 10,000 times to get a distribution of the K-S test statistic for a $\chi^{2}$ distribution with $2\times13$ d.o.f. for a given sample size of $N_{\rm sample}$ (=130). This bootstrap method accounts for the size of the sample. The resulting distribution is shown as a cyan histogram in the Extended Data Fig. 7 (c). It is clear that \srcs’s observed K-S test statistic is lower than the median value of the distribution. This suggests that noise powers are consistent with the expected $\chi^{2}$ distribution, i.e., \srcs’s noise powers in the vicinity of the QPO are consistent with being white.
\par
We also investigated the goodness-of-fit with the Anderson-Darling test. We computed the statistic distribution using the same bootstrap technique described above. This resulting distribution is shown in the Extended Data Fig. 7 (d). Again, the value of Anderson-Darling’s test statistic indicates that the observed noise powers of J2150-0551’s PDS are consistent with being $\chi^{2}$ distributed.
To validate the results, we also used the same methods to test the noise properties of the PDS by
extracting the PN light curve only, using its minimum time resolution of 73.4 ms.
This enables the PDS extending to higher frequencies (6.8 Hz), hence is better to evaluate the noise type.
We found consistent results that the powers' distribution in the PDS is dominated by white noise.

\par
To summarize, all the statistical tests lead to the same conclusion that the noise powers above 7.65$\times10^{-4}$ Hz are consistent with being white.
The above rigorous analysis of the noise properties of the PDS continuum will
enable to accurately assess the QPO significance.


{\bf \noindent Statistical Significance under white noise:}
For the white noise distribution of the underlying continuum in the PDS,
the significance of the QPO at $\sim$11.88 mHz can be analytically estimated using the $\chi^{2}$ distribution.
We found that the probability of observing a power value larger than the measured peak value at the QPO frequency ($\xi_{peak}=$ 6.14) is
P$(>$$13\xi_{peak}; \rm d.o.f)=2.2\times10^{-7}$$=p_{\rm 1}$.
By taking into account the search for all frequency bins over the range of interest ($N= 252$),
the global significance is 1-$p_{\rm 1}\times$N= 99.9944\%, equivalent to  4.03$\sigma$ assuming a normal distribution.

{\bf \noindent Statistical Significance with Monte-Carlo simulations:}
The above assessment of noise distribution in the PDS is
only qualitative. It is still possible that a weak or unknown red
noise component is present in the data. To estimate the QPO
significance by adequately accounting for the underlying noise,
we then used the power-law plus constant model, $ P(f) = Nf^{-\alpha} + C $,
to fit the unbinned PDS but excluded the 15 frequency bins near the QPO frequency, where N is the normalization factor, $ \alpha $ is the power-law index,
and C is a constant indicating the Poisson noise level. The powerlaw component was used to account for the possible
effect of weak red noise. We used the maximum likelihood estimation method described in ref. \citep{Vaughan2010, Barret2012, Lin2013} to obtain the best-fit parameters. The best-fitting powerlaw normalization and index value were only poorly constrained to be 2.5$^{+2.7}_{-2.5}\times10^{-6}$ and 0.35$^{+0.81}_{-0.35}$, respectively. In order to fully account for the statistical uncertainty of the model and better ascertain the significance of the QPO detection
under the assumption that the underlying noise is red,
we used the Markov Chain Monte Carlo (MCMC) method to map the multi-dimensional distribution of the model parameters N and C by fixing the power-law index at various values between 0.3 and 1.7 ($\sim$90\% confidence interval, see also ref. \citep{Pasham2019}).
We used the emcee sampler with a Gaussian likelihood function\citep{Foreman-Mackey2013}, employing 20 walkers and 10$^{6}$ steps. The first 2000 steps were discarded for convergence, and a thinning factor 20 was applied to reduce autocorrelation. A flat prior was applied to both parameters: 0 < N < 1 and 1.5 < C < 2.5. Then approximately 10$^{6}$ sets of parameters are obtained for different $\alpha$ fixed 0.3, 0.7, 1.0, 1.4 and 1.7, respectively. The corresponding posterior distribution of the model parameters is shown in Extended Data Fig. 8. 
Subsequently, we randomly drew the values of parameters N and C for each simulation, which can generate a simulated light curve for a given PDS shape with $\alpha$ fixed\citep{Timmer1995}.
Each simulated light curve was resampled to have the same duration, mean count rate, and variance as the real light curve (GTI2 in XMM2). To eliminate the possible effect of red noise leakage at the edge of the light curve, we choose a length of $5\times 17000$ s, which is five times longer than the real data.  Then, we intercepted the middle 17 ks as the final light curve for simulations.
We repeated this process and obtained 500,000 simulated light curves. The same power spectral analysis on these simulated light curves was performed as we did on the real data, producing 500,000 simulated PDS.

For each fixed red-noise slope, with its corresponding 500,000 simulated PDS, we computed the significance level of a QPO peak by scanning all frequency bins below 0.19 Hz, the highest
frequency in the PDS. We averaged the power of the 500,000 PDS at each frequency bin and obtained the $ 	\left \langle P_{f_{i}} \right \rangle$ where $i$ is $i^{\rm th}$ number of bins ($i=1, \dots, n$).
For each simulated PDS, we divided the power value at each frequency by the average value $ 	\left \langle P_{f_{i}} \right \rangle$ at the corresponding frequency, and recorded the largest value as $\xi_{max}$, yielding 500,000 $\xi_{max}$ from all simulated PDS. Averaging the power distribution can eliminate the possible effect of red noise \citep{Pasham2019}.
Using the 500,000 values of $\xi_{max}$, we calculated the cumulative
distribution of probability to exceed a given $\xi_{max}$ (Extended Data Fig. 9).
By comparing with the observed value of $\xi_{max,obs}$, we
obtained the global confidence level of the QPO at 11.88 mHz for each red-noise slope. The simulation results are summarized in Table S2.
With the above rigorous approach of light curve simulations, which fully accounts for
the statistical uncertainty of the model,
we found that the observed QPO is significant at a level greater than 3.51$\sigma$.


{\bf \noindent Ruling out instrumental or background origin:}
We check the possibility that the QPO signal is caused by background fluctuations by extracting the PDS
of a blank sky field during the same observing epoch as \src where
the QPO is detected (XMM2).
The corresponding Leahy-normalized PDS from three individual light curve segments are shown in Extended Data Fig. 10 (b),
which are dominated by noise.
In addition, we also extracted the PDS of two other X-ray sources, which have been observed in the same
field and epoch, with PN count rates comparable to \srcs.
Extended Data Fig. 10 (c) and (d) show the resulting PDS and no strong QPO feature at 11.88 mHz is seen in any PDS.

{\bf \noindent 4. X-ray spectral analysis}

 Ref. \cite{Wen2021} has simultaneously fitted the five-epoch X-ray spectra of J2150-0551 by using a modified version of the relativistic slim disk model to estimate the black hole mass and spin of J2150-0551. However, the value of the spectral hardening factor $f_{c}$ was not well studied during the super-Eddington accretion phase, and therefore, the influence of $f_{c}$ on parameter estimation remains uncertain. Furthermore, the slim disk model may encounter challenges in accurately describing disks with extremely high accretion rates, as the vertical hydrostatic equilibrium equation may be compromised when the disk becomes excessively thick. Additionally, in the case of early TDE disks, alignment with the equatorial plane of the black hole may not occur, rendering the stationary disk model inadequate for describing such configurations. Consequently, we have opted to exclude the initial two spectra from our refitting process\cite{Davis2019, Wen2020}.

 We choose to fit the PN spectrum between 0.2-3.0 keV from the XMM2 observation in which the QPO is detected, using the spectral fitting software \texttt{XSPEC} (Version 12.13.1). We grouped the spectrum to have at least 15 counts in each bin so as to adopt the $\chi^{2}$ statistic for the spectral fits. All statistical errors of spectral model parameters will be reported at the 90\% confidence level for one parameter of interest ($\Delta$$\chi^{2}$ = 2.706). We fitted the X-ray spectrum using the slim disk model modified by two absorption components (phabs*zphabs*slimd) \cite{Wen2022}. The first absorption component was fixed at $N_{\rm H}$ = 2.6$\times10^{20}$ cm$^{-2}$ to account for Galactic absorption \cite{Lin2018}, while the second was allowed to float to account for the intrinsic absorption of J2150-0551. The model perfectly describes the spectrum, yielding a $\chi^{2}$/d.o.f. = 209.48/209. In this case, the best-fitting model results in an intrinsic absorption $N_{\rm H}$ = 4.0$^{+0.9}_{-0.7}$$\times10^{20}$ cm$^{-2}$, accretion rate $\dot{m}$ = 1.6$^{+0.6}_{-0.5}$, observer’s inclination angle $\theta$ = 15$^{+13}_{-8}$ degree, black hole mass $M_{\rm BH}$ = 1.3$^{+2.5}_{-0.6}$$\times10^{4}$ \msun. The XMM2 spectrum, along with the best-fitting model, is shown in Extended Data Fig. 11. Based on the best-fitting model, we constructed two-parameter, joint-confidence contours of the BH mass versus dimensionless spin, and the results are shown in Figure 3.
After a test, we found that if we simultaneously fitted the X-ray spectra by combining the data from the other two XMM observations at lower flux levels, we could obtain results consistent with that of fitting to the XMM2 data alone.
Note that the slim disk model provides a more reasonable constraint on 
the black hole mass in the context of TDEs, which 
is not at odds with that obtained from the spectral fittings with standard thin disk if a lower value of spin is assumed \cite{Lin2018}.


\end{methods}

 \begin{addendum}
   \item [Data availability]
	
	   {Source data for the observations taken with \xmm are available through
	the HEASARC online archive services (https://heasarc.gsfc.nasa.gov/docs/archive.html).
	 The authors can provide other data that support the findings of this study upon request.
	   }
   \end{addendum}

  \begin{addendum}
	     \item [Code availability]
		     {{The Science Analysis Software used to reduce the \xmm data is publicly available at
		          https://www.cosmos.esa.int/web/xmm-newton/download-an.
	       The python code to conduct the power spectral analysis is available
	       at https://github.com/StingraySoftware/stingray \citep{Huppenkothen2019}.
             The other codes that support the plots within this article are available from the
     authors upon reasonable request.}}
  \end{addendum}

\noindent{\bfseries References}\setlength{\parskip}{12pt}%

\input{main_v3.bbl}



\begin{addendum} 
 \item[Correspondence] Correspondence and requests for materials should be addressed to Xinwen Shu (xwshu@ahnu.edu.cn).

 \item [Acknowledgments]
 This research made use of the HEASARC online data archive services, supported by
NASA/GSFC. We thank the \xmm observatory
for making the data available.
We thank S. X. Wen for providing the slim disk model for X-ray spectral fittings,
and for helpful discussions on interpreting the spectral fitting results.

 This work is supported by National Natural
Science Foundation of China through grant Nos. 12192220 and 12192221,
National Key Research and Development Program of China
(2022SKA0130102), the China Manned Spaced Project (CMS-CSST-2021-A06) and the Postdoctoral Fellowship Program of CPSF under Grant Number GZC20232699.

\item[Author Contributions]
X.W.S. conceived the project and authored the majority of the text. 
W.J.Z. led the data reduction, performed X-ray timing and spectral analysis, and wrote the draft. %
L.M.S. contributed to the timing analysis and commented on the manuscript.
T.G.W. and R.F.S. contributed to the overall interpretation of the results.
All the authors joined the discussion at all stages.

 \item[Competing Interests] The authors declare no competing interests.
\end{addendum}
\newpage

\begin{table*}
  \centering
  \caption{{\bf Extended Data Tab. 1: XMM-Newton data used in power spectral analysis}}
  \begin{tabular}{ccccccc}
    \hline
    \hline
Obs No. & ObsID & Obs date & GTI & Exposure  & PN count rates (0.2-2 keV) & $f_{\rm QPO}$   \\
&&&& ks & cts s$^{-1}$ & mHz \\
    \hline
XMM1 & 0404190101& 2006-05-05 & GTI1 & 8 &  0.29\ppm0.07& $-$ \\
  \hline
\multirow{3}*{XMM2} &\multirow{3}*{0603590101} & \multirow{3}*{2009-06-07} & GTI2 &17& 0.10\ppm0.03&11.88  \\
&&& GTI3 &21& 0.11\ppm0.03&$-$ \\
&&& GTI4 &20& 0.11\ppm0.03&$-$ \\
  \hline
XMM3&0823360101 & 2018-05-24 & GTI5 & 57&  0.02\ppm 0.01&$-$ \\

\hline

\hline

  \end{tabular}

\begin{tablenotes}
\item{\textbf{\large{Note}}-List of XMM–Newton observations of J2150-0551. Five GTIs have been chosen to construct PDS, and only GTI2 shows a QPO signal.
The PN count rates encompass both source and background counts. }
\end{tablenotes}
\end{table*}

\begin{table*}
  \centering
  \caption{{\bf Extended Data Tab. 2: Statistical significance under red noise by fixing the red-noise slope at different values}}
  \begin{tabular}{ccc}
    \hline
    \hline
Red-noise slope ($\alpha$) &  number of ($\xi_{max}<\xi_{max,obs}$)\textsuperscript{†} & Significance  \\
\hline
 0.3&  499776 & 3.51$\sigma$ \\
\hline
0.7&  499815 & 3.56$\sigma$ \\
\hline
1.0&  499817 & 3.56$\sigma$ \\
\hline
1.4&  499818 & 3.56$\sigma$ \\
\hline
1.7&  499825 & 3.58$\sigma$ \\

\hline

\hline

  \end{tabular}

\begin{tablenotes}
\item{\textbf{\large{Note}}-\textsuperscript{†}Number of simulations with $\xi_{max}<\xi_{max,obs}$ out of 500,000 simulations.}
\end{tablenotes}
\end{table*}

\begin{figure}[ht]
\centering
\includegraphics[width=\linewidth]{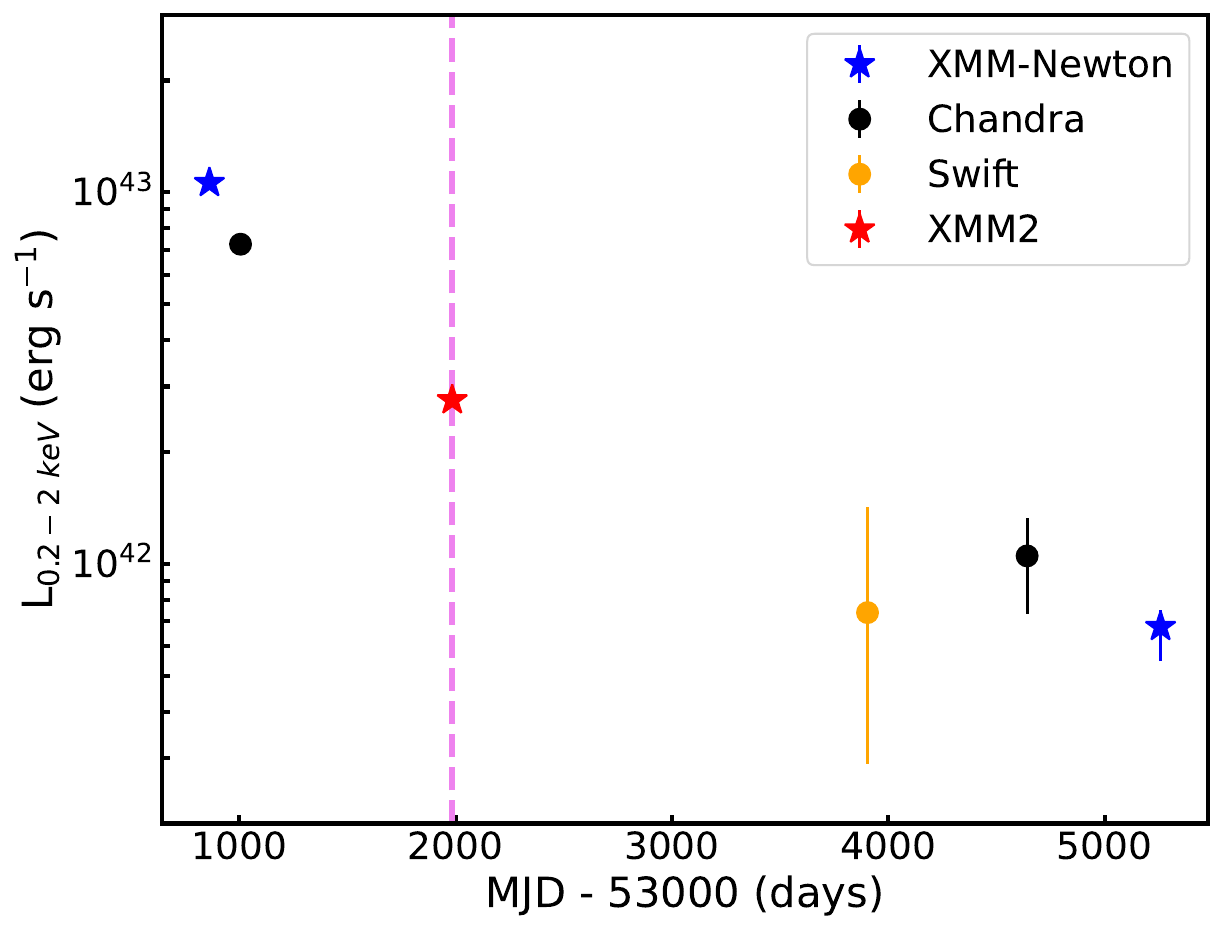}
\caption{\textbf{Extended Data Fig. 1: X-ray luminosity evolution of J2150-0551 observed with \xmmp, \swiftp,
and Chandra.}
The vertical dashed line shows the observation (XMM2) in which the QPO is detected.
The data are taken from ref.\citep{Lin2020}.
}
\end{figure}

\begin{figure}[ht]
\centering
\includegraphics[width=\linewidth]{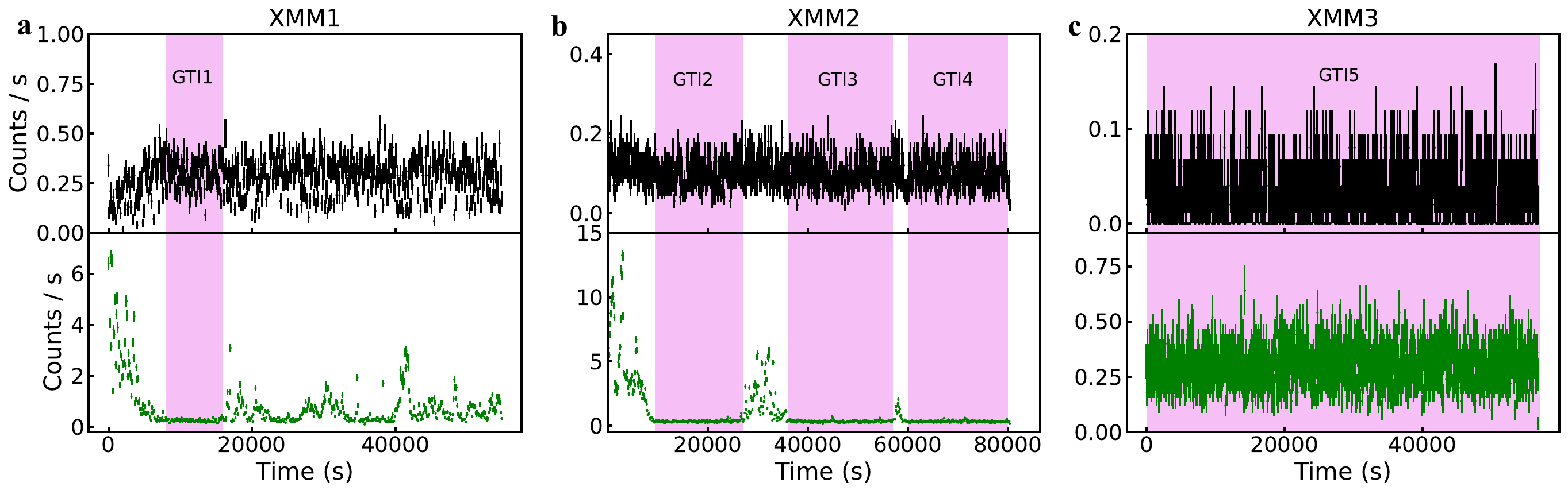}
\caption{\textbf{Extended Data Fig. 2: XMM-Newton/PN 0.2–2 keV light curves for J2150-0551 (black) and background (green)}. After filtering the high background periods, the light curve segments used for constructing the PDS are highlighted by shaded purple rectangles.
}
\end{figure}

\begin{figure}[ht]
\centering
\includegraphics[width=0.5\linewidth]{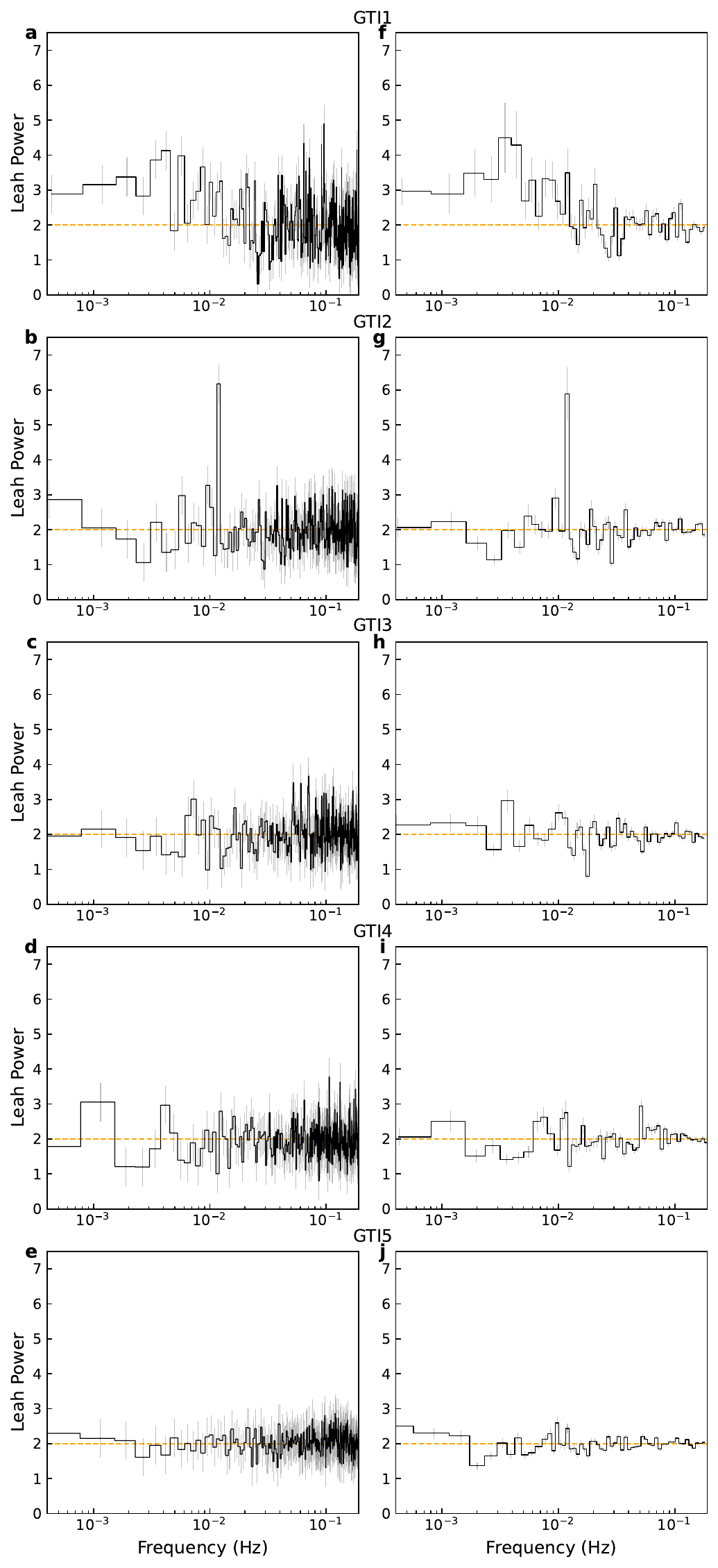}
\caption{{\bf Extended Data Fig. 3: Power spectra of all the five individual GTIs.}  a-e: Linear rebinning PDS with a frequency step size of $\sim$7.6$\times10^{-4}$ Hz for GTI1 - GTI5. f-j: Logarithmic rebinning of the PDS for GTI1 - GTI5 with a rebinning factor of 1.07. This clearly shows that the normalized PDS at higher frequencies is consistent with a value of 2.0, i.e., dominated by white noise.}

\end{figure}

\begin{figure}[ht]
\centering
\includegraphics[width=\linewidth]{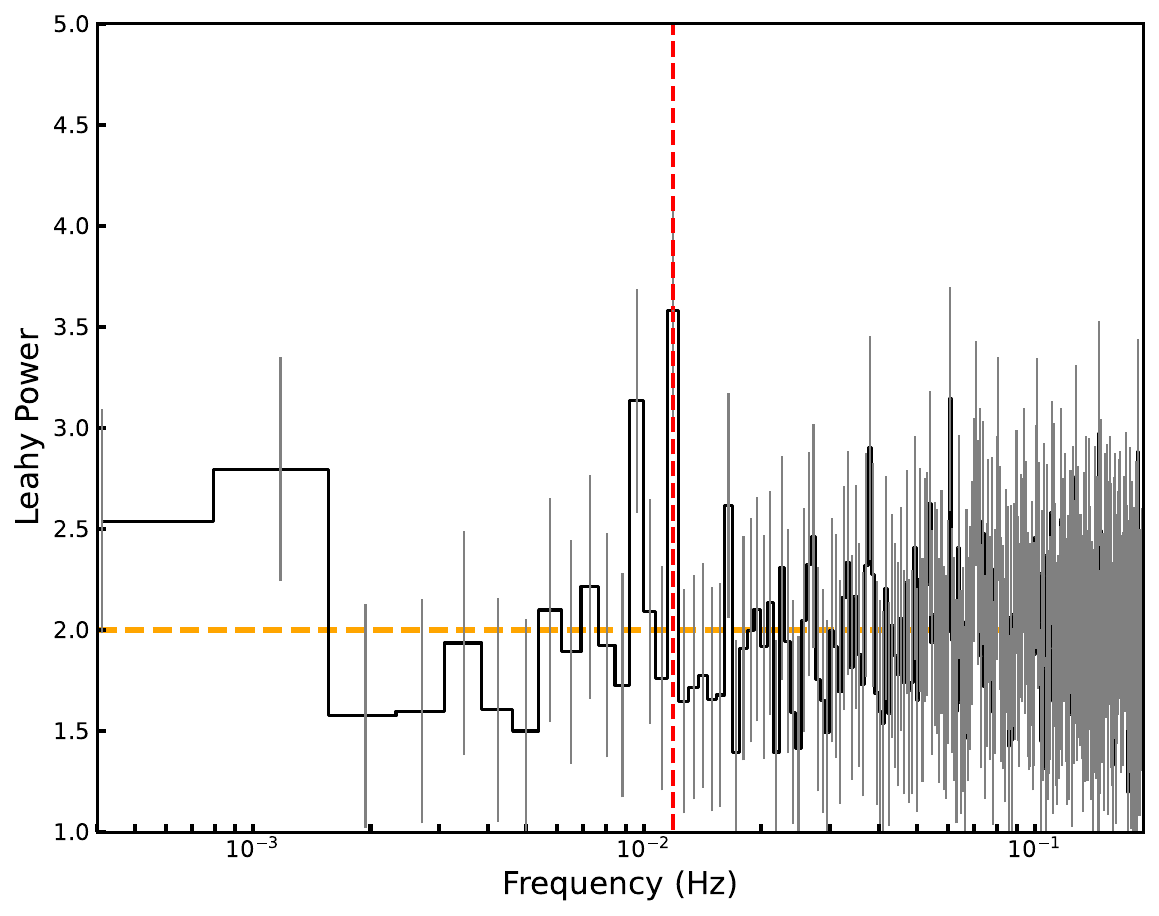}
\caption{\textbf{Extended Data Fig. 4: Averaged PDS by combining the PDS of GTI2-GTI4 with a uniformly selected exposure time of 17 ks.} The red dashed line shows the frequency of QPO at 11.88 mHz.
}
\end{figure}

\begin{figure}[ht!]
\centering
\includegraphics[width=\linewidth]{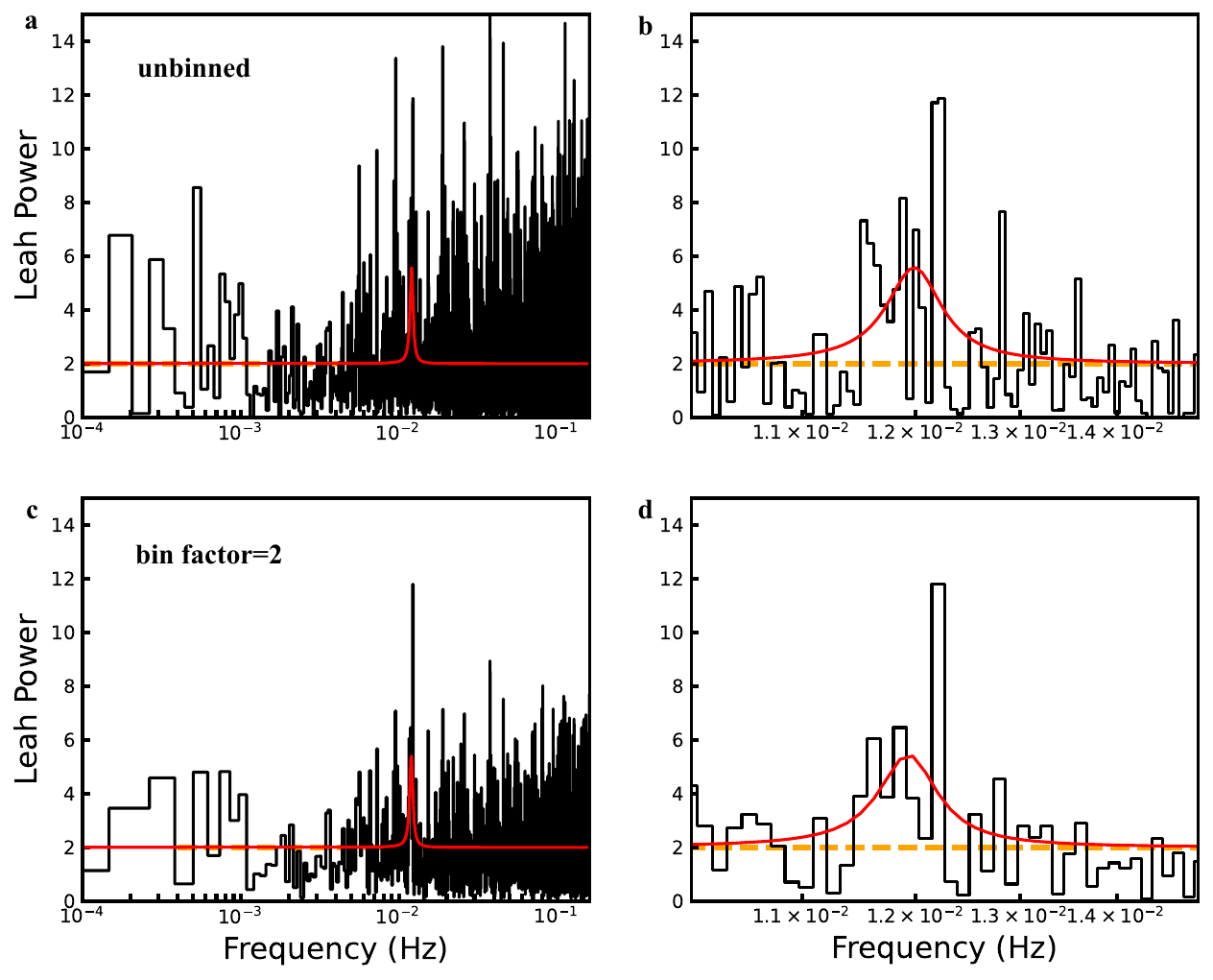}
\caption{\textbf{Extended Data Fig. 5: PDS of GTI2 fitted by a powerlaw+constant plus a Lorentzian model.} The powerlaw+constant model accounts for the continuum, while the Lorentzian model is used to fit the QPO signal. Top left panel: unbinned PDS along with the best-fit model (red curve).
Top right panel: a zoomed-in view of the fitting result covering the QPO frequency.
Bottom two panels are the same as Top, but for the PDS averaged in frequency by a factor of 2.
In either case, the FWHM of Lorentzian model is found to be $\sim$7.6$\times10^{-4}$ Hz, which is $\sim$13 times the minimum frequency resolution ($5.88\times10^{-5}$ Hz) of the unbinned PDS.
}
\end{figure}


\begin{figure}[ht]
\centering
\includegraphics[width=\linewidth]{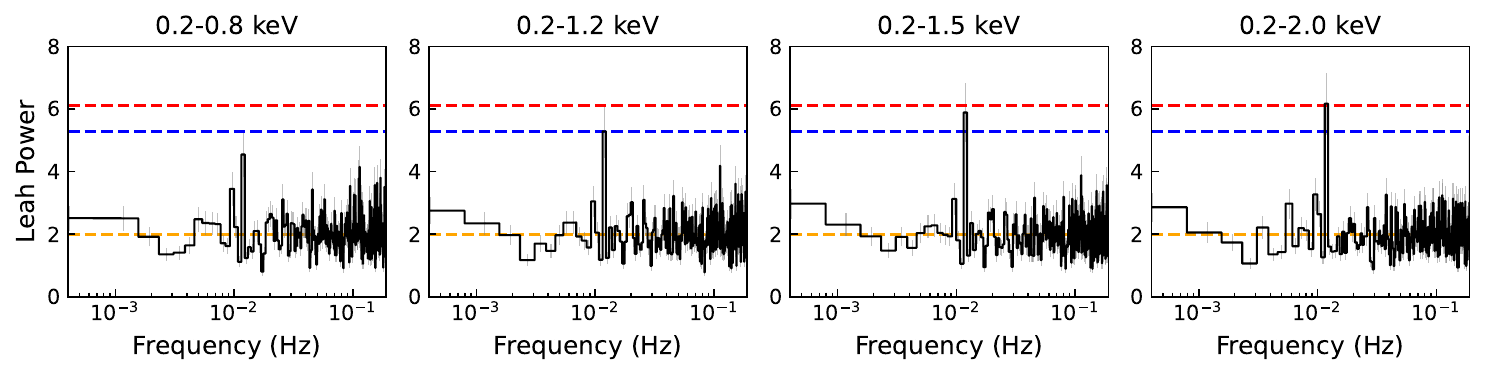}
\caption{\textbf{Extended Data Fig. 6: The PDS of GTI2 light curve at four different energy bands.}  {\bf a} The PDS of 0.2-0.8 keV light curve. {\bf b} The PDS of 0.2-1.2 keV light curve. {\bf  c} The PDS of 0.2-1.5 keV light curve. {\bf d} The PDS of 0.2-2.0 keV light curve. The blue and red dashed lines represent the 3$\sigma$ and 4$\sigma$ confidence levels, respectively. This demonstrates that the entire 0.2–2 keV energy band contains the QPO component.}

\end{figure}

\begin{figure}[ht]
\centering
\includegraphics[width=\linewidth]{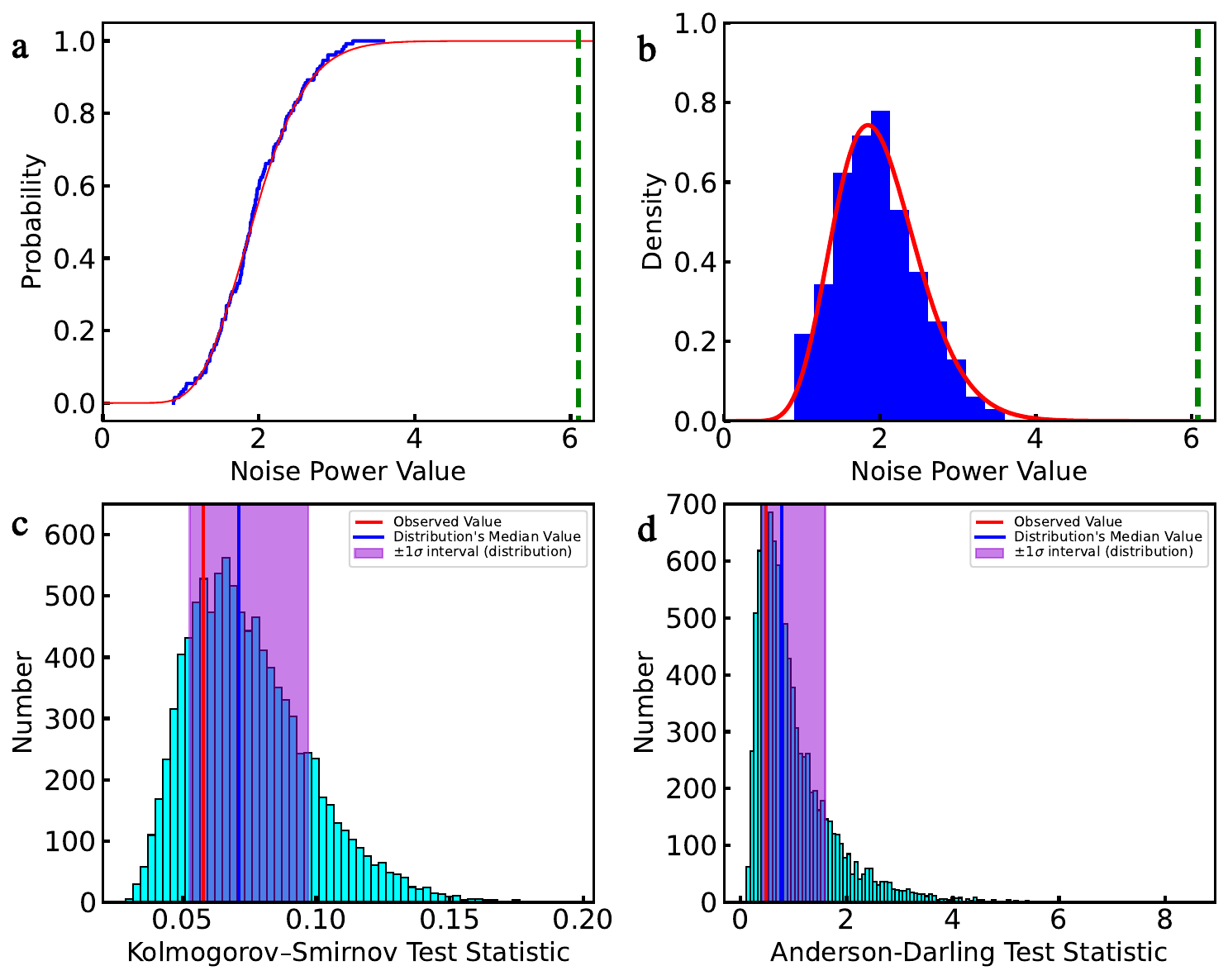}
\caption{{\bf Extended Data Fig. 7:}
{\bf Tests for white noise in the PDS continuum by combining PN and MOS light curves with a time resolution of 2.6s.} {\bf a}: the CDF of PDS noise powers in the range 7.65$\times10^{-4}-0.19$ Hz (ignoring three frequency bins centered on the QPO's centroid frequency). The blue histogram is the data while the red curve is the expected $\chi^{2}$ distribution for white noise. The CDF tracks the expected CDF over the range of observed power values. The green dashed line marks the observed power value of the QPO. {\bf b}: the probability density function of the observed noise powers compared with the expected $\chi^{2}$ distribution for white noise. {\bf c}: the distribution of the K-S statistic derived from CDFs sampled from a $\chi^{2}$ distribution with 26 d.o.f. The red line represents the K-S test statistic value for the observed data, which is lower than the median of the distribution (blue line) obtained from bootstrap simulations. {\bf d}: the same as {\bf c} but using an Anderson-Darling test statistic. Both the test statistic values are consistent with a $\chi^{2}$ distribution and suggest that the PDS continuum is
consistent with white noise.
}
\end{figure}

\begin{figure}[ht]
\centering
\includegraphics[width=0.75\linewidth]{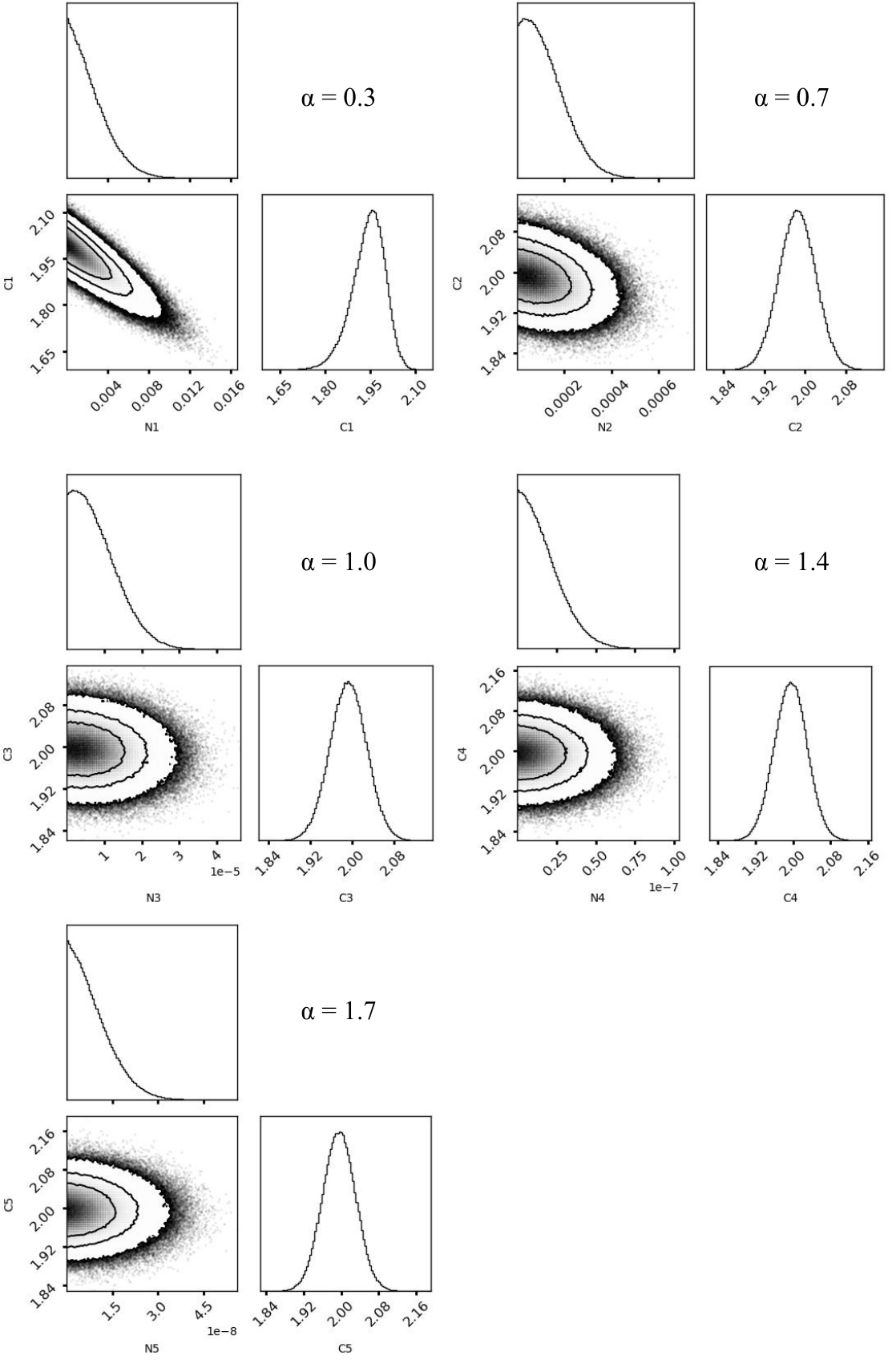}
\caption{{\bf Extended Data Fig. 8:}
Posterior distribution of parameters $N$ and $C$, obtained by fitting a powerlaw + constant model ($ P(f) = Nf^{-\alpha} + C $) to the unbinned PDS of the PN+MOS lightcurve, where the powerlaw index $\alpha$ was fixed at five different values. Black contour lines enclose the confidence intervals corresponding to 1$\sigma$, 2$\sigma$ and 3$\sigma$ assuming Gaussian distribution.
}
\end{figure}

\begin{figure}[ht]
\centering
\includegraphics[width=0.95\linewidth]{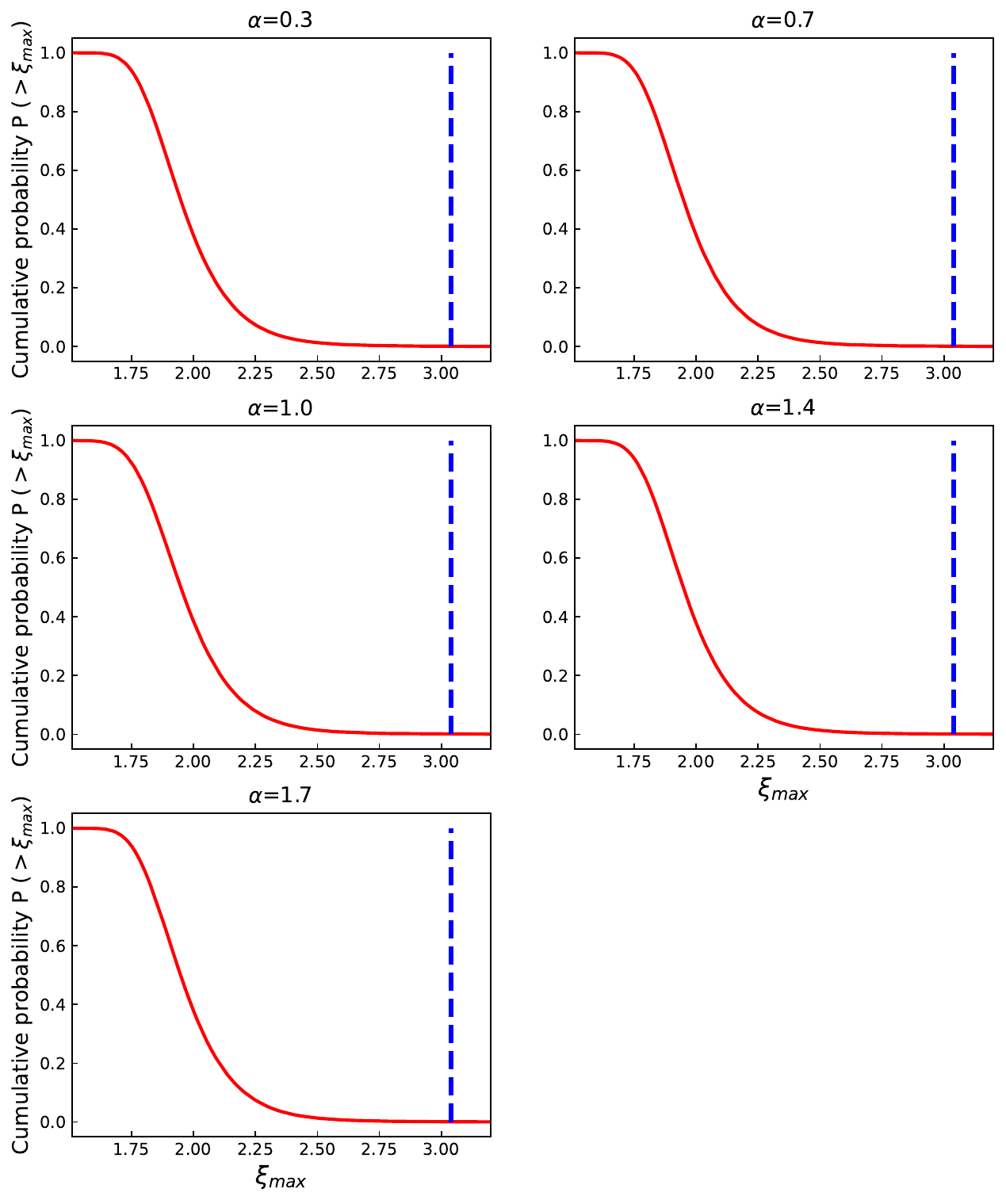}
\caption{{\bf Extended Data Fig. 9:
Cumulative distribution of the maximum noise power P ($>$$\xi_{max}$) from Monte Carlo simulations by assuming different red noise slope.} Different panels show the results from Monte Carlo simulations by fixing the red noise index at 0.3, 0.7, 1.0, 1.4 and 1.7, respectively. In the simulations, the parameter values of N and C are randomly selected from the posterior distribution (Extended Data Fig. 8).}
\end{figure}

\begin{figure}[ht]
\centering
\includegraphics[width=0.85\linewidth]{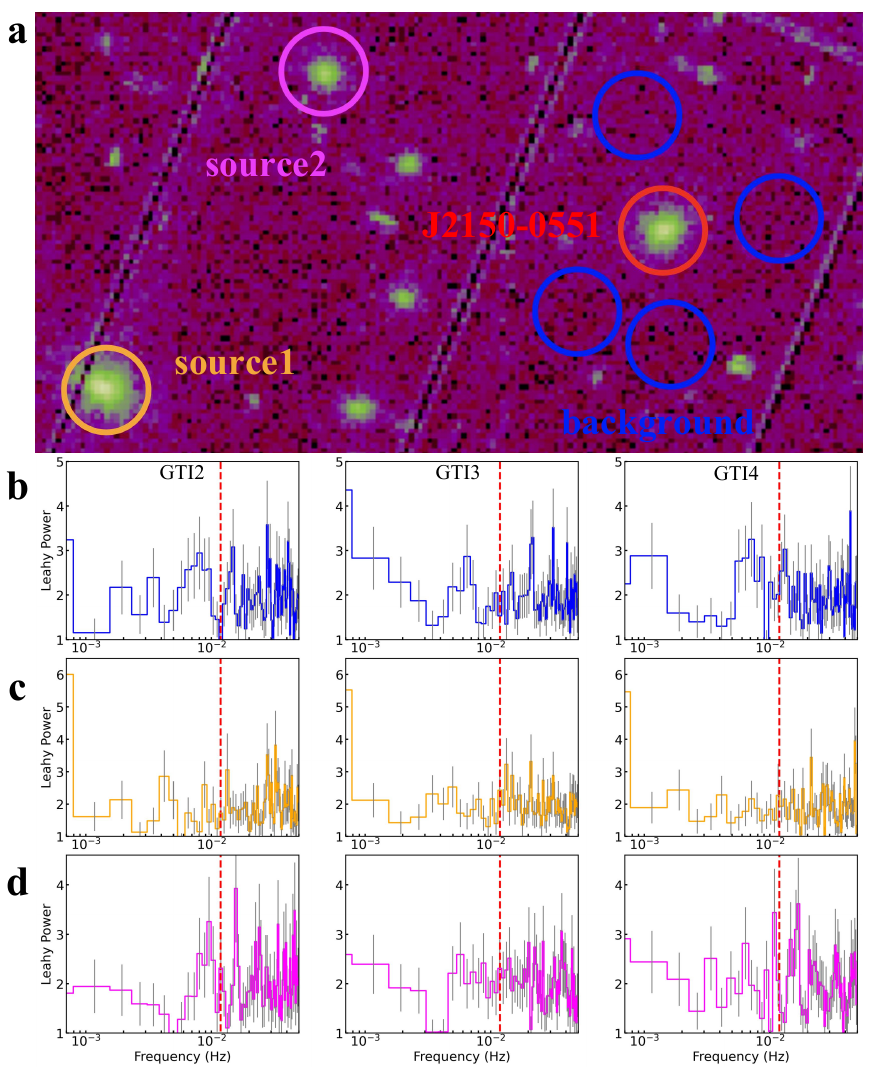}
\caption{\textbf{Extended Data Fig. 10: Power spectra of sky background and other X-ray sources in XMM2}.
Panel (a) shows the XMM2/PN image around J2150-0551. We show the circular extraction region for the X-ray emission of J2150-0551 (red) and background regions (blue). The light curves from other two nearby sources are also extracted for comparison, source1 (orange) and source2 (magenta), respectively.  Panel (b) shows the power spectra of background in three individual light curve segments (GTI2-GTI4), respectively.  Panels (c) and (d) show the power spectra of source1 and source2 in GTI2-GTI4, respectively. Red dashed lines represent the QPO frequency at 11.88 mHz. }
\end{figure}

\begin{figure}[ht]
\centering
\includegraphics[width=\linewidth]{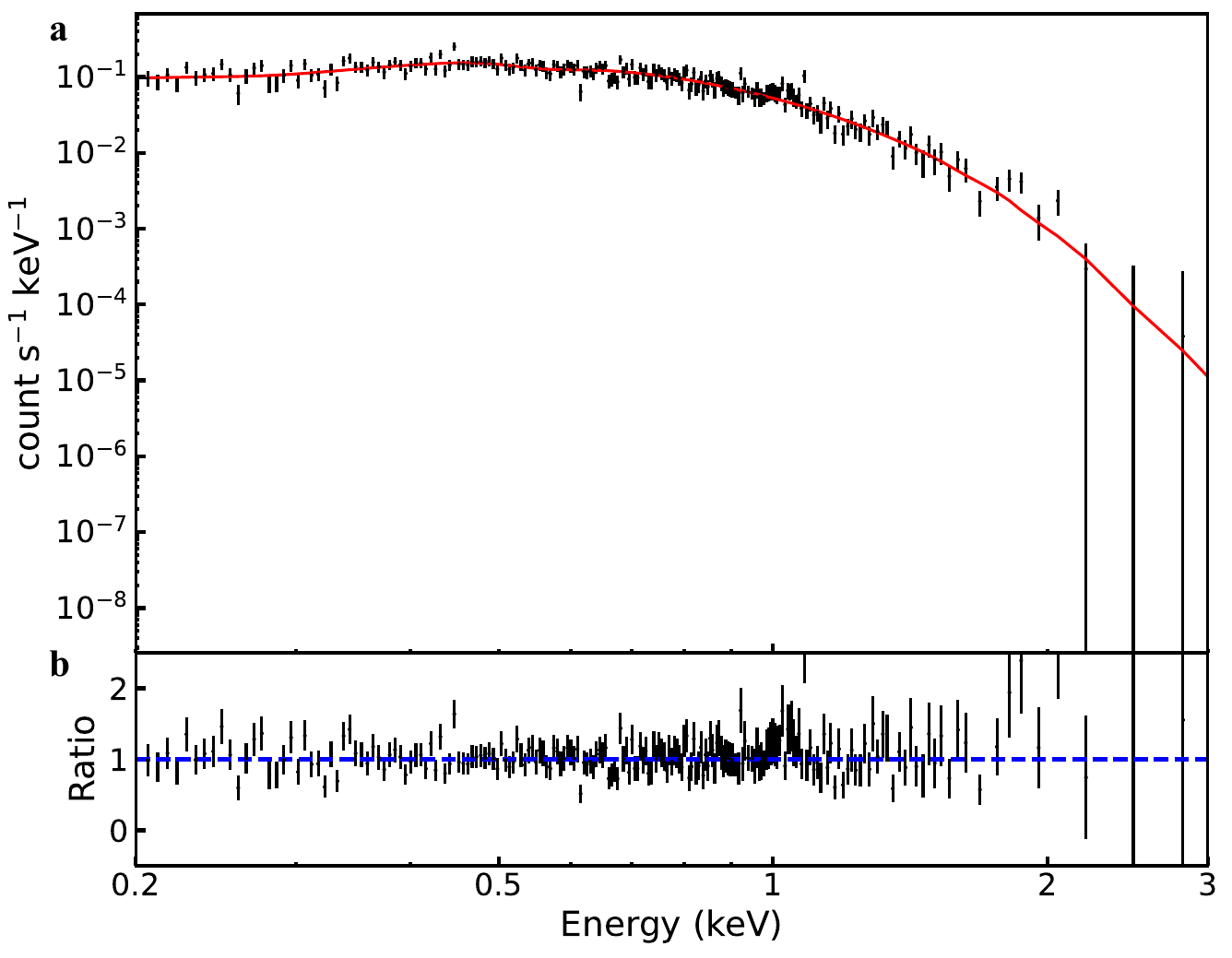}
\caption{{\bf Extended Data Fig. 11: Fitting results to the X-ray spectrum of J2150-0551 from XMM2 observation in which the QPO is detected.} {\bf (a)}: XMM2 PN spectrum along with the best-fitting slimdisk model (red curve). The corresponding data to model ratios are shown in {\bf (b)}. Error bars represent 1$\sigma$ uncertainties.
}
\end{figure}



\end{document}

%% file: preamble.tex
\usepackage{lineno}

\usepackage{setspace}
\usepackage{amsfonts}
\usepackage{amssymb}
\usepackage{amsmath}

\usepackage{float}
\usepackage{placeins}
\usepackage{graphicx}

\usepackage[export]{adjustbox}

\usepackage[nomarkers,nolists,figuresonly]{endfloat}

\usepackage[normal,normal,bf,labelformat=empty]{caption}

\newcounter{mybodyfigure}
\newcounter{myedfigure}

\newcommand{\beginbodyfigures}{\renewcommand{\thefigure}{{\themybodyfigure}}}

\newcommand{\beginedfigures}{\renewcommand{\thefigure}{{\themyedfigure}}}

\newcommand{\stepbodyfigure}{\refstepcounter{mybodyfigure}}

\makeatletter
\g@addto@macro\caption@prepareslc{%
  \renewcommand{\stepbodyfigure}{\caption@l@stepcounter{mybodyfigure}}}
\makeatother

\newcommand{\stepedfigure}{\refstepcounter{myedfigure}}

\makeatletter
\g@addto@macro\caption@prepareslc{%
  \renewcommand{\stepedfigure}{\caption@l@stepcounter{myedfigure}}}
\makeatother

\usepackage[super,comma,sort&compress]{natbib}

\usepackage{hyperref}
\usepackage{verbatim}

\input{preamble_optional.tex}

%% file: preamble_optional.tex
\usepackage{ulem} 
\usepackage{rotating} 
\usepackage{soul}

\usepackage{color} 

\usepackage[colorinlistoftodos]{todonotes}

\makeatletter
\newsavebox\myboxA
\newsavebox\myboxB
\newlength\mylenA

\newcommand*\xoverline[2][0.75]{%
    \sbox{\myboxA}{$\m@th#2$}%
    \setbox\myboxB\null
    \ht\myboxB=\ht\myboxA%
    \dp\myboxB=\dp\myboxA%
    \wd\myboxB=#1\wd\myboxA
    \sbox\myboxB{$\m@th\overline{\copy\myboxB}$}
    \setlength\mylenA{\the\wd\myboxA}
    \addtolength\mylenA{-\the\wd\myboxB}%
    \ifdim\wd\myboxB<\wd\myboxA%
       \rlap{\hskip 0.5\mylenA\usebox\myboxB}{\usebox\myboxA}%
    \else
        \hskip -0.5\mylenA\rlap{\usebox\myboxA}{\hskip 0.5\mylenA\usebox\myboxB}%
    \fi}
\makeatother
